\documentstyle[epsfig,a4,12pt,rotating]{article}

\newcommand{\Date}      {7 July, 1999}

\def\mrm{\mathrm}

\newcommand{\epm} {\mbox{$\mathrm{e}^+ \mathrm{e}^-$}}

\newcommand{\Hosm}{\mbox{$\mathrm{H}^{0}_{\mathrm{SM}}$}}

\newcommand{\ZZ}{\mbox{$\mathrm{Z}^{0}{\mathrm{Z}^{0}}^{(*)}$}}

\newcommand{\leplep}{\mbox{$\ell^+\ell^-$}}

\newcommand{\lnu}{\mbox{$\ell\bar\nu_\ell$}}

\newcommand{\evisn}{\mbox{$\protect E_{\mathrm vis}$/$\protect \sqrt{s}$}}
\newcommand{\rvis}{\mbox{$R_{\mrm{vis}}$}}

\newcommand{\wwln}{\mbox{$\mrm{W}^{+}\mrm{W}^{-}
            \ra\mathrm{q}{\bar{\mathrm{q}}}\lnu$}}

\newcommand{\mstopL}{\mbox{$m_{\tilde{\mathrm{t}}_{\mathrm{L}}}$}}
\newcommand{\mstopR}{\mbox{$m_{\tilde{\mathrm{t}}_{\mathrm{R}}}$}}

\newcommand{\Gc}{\mbox{$\mathrm{GeV}$}}

\newcommand{\Dbi}{\mbox{${\cal B}_i$}}

\newcommand{\ee}{\mbox{${\mathrm{e}}^+ {\mathrm{e}}^-$}}
\newcommand{\tautau}{\mbox{$\tau^+\tau^-$}}
\newcommand{\mm}{\mbox{$\mu^+\mu^-$}}

\newcommand{\qq}         {\mbox{$\mathrm{q}\bar{\mathrm{q}}$}}
\newcommand{\bb}         {\mbox{$\mathrm{b}\bar{\mathrm{b}}$}}
\newcommand{\ff}         {\mbox{$\mathrm{f}\bar{\mathrm{f}}$}}
\newcommand{\nunu}       {\mbox{$\nu\bar{\nu}$}}
\newcommand{\mZ}         {\mbox{$m_{\mathrm{Z}}$}}

\newcommand{\mh}         {\mbox{$m_{\mathrm{h}}$}}
\newcommand{\mA}         {\mbox{$m_{\mathrm{A}}$}}

\newcommand {\Ho}        {\mbox{$\mathrm{H}^{0}$}}
\newcommand {\Ao}        {\mbox{$\mathrm{A}^{0}$}}
\newcommand {\ho}        {\mbox{$\mathrm{h}^{0}$}}
\newcommand {\Zo}        {\mbox{$\mathrm{Z}^{0}$}}
\newcommand{\Zs}         {\mbox{${\mathrm{Z}}^{*}$}}
\newcommand{\Zgs}        {\mbox{$\mathrm{(Z/\gamma)}^{*}$}}

\newcommand{\nn}{\mbox{$\nu \bar{\nu}$}}

\newcommand{\WW}         {\mbox{$\mathrm{W}^+\mathrm{W}^-$}}
\newcommand{\pb}         {\mbox{$\mathrm{pb}^{-1}$}}
\newcommand{\tanb}       {\mbox{$\tan\!\beta$}}
\def\mrm       {\mathrm}

\newcommand{\sqrts}     {\sqrt{s}}
\newcommand{\ZPhysC}[1]    {Z. Phys. {\bf C#1}}

\newcommand{\PhysLettB}[1] {Phys. Lett. {\bf B#1}}

\def\etal{\mbox{{\it et al.}}}

\newcommand{\ra}        {\mbox{$\rightarrow$}}   
\begin{document}
\begin{titlepage}
\centerline{\Large EUROPEAN LABORATORY FOR PARTICLE PHYSICS}
\bigskip
\begin{flushright}
      CERN-EP/99-096 \\ 
      \Date
\end{flushright}
\bigskip\bigskip\bigskip

\begin{center}{\LARGE\bf Search for Neutral Higgs Bosons \\
in {\boldmath \ee\unboldmath} Collisions at 
{\boldmath $\sqrt s \approx$ \unboldmath}189~GeV }
\end{center}
 \bigskip
\begin{center}{\LARGE The OPAL Collaboration}
\end{center}
\bigskip
\bigskip
\begin{center}{\large  Abstract}\end{center}
A search for neutral Higgs bosons has been performed 
with the OPAL detector at LEP, using approximately
170 pb$^{-1}$ of \epm\ collision data
collected at $\sqrt{s} \approx 189$~GeV.
Searches have been performed for the Standard Model (SM) process
\ee\ra\Ho\Zo\  and the MSSM processes $\ee\ra\ho\Zo$, $\Ao\ho$.
The searches are sensitive to the
\bb\ and \tautau\ decay modes of the Higgs bosons, and also to 
the MSSM decay mode
\ho\ra\Ao\Ao.  OPAL search results at
lower centre-of-mass energies have been incorporated 
in the limits we set, which are valid at the 95\% confidence level.
For the SM Higgs boson, we obtain a lower mass bound of 91.0 GeV.
In the MSSM, our limits are
\mh$>74.8$~GeV and \mA$>76.5$~GeV, assuming
\tanb$>$1, that the mixing of the scalar top quarks is either zero or maximal,
and that the soft SUSY-breaking masses are 1~TeV.  For the case of zero
scalar top mixing, we exclude values of $\tan\beta$ between 0.72 and 2.19.

\bigskip\bigskip

\begin{center}
  {\large (Submitted to European Physical Journal C)}
\end{center}
\end{titlepage}

\begin{center}{\Large        The OPAL Collaboration
}\end{center}\bigskip
\begin{center}{
G.\thinspace Abbiendi$^{  2}$,
K.\thinspace Ackerstaff$^{  8}$,
G.\thinspace Alexander$^{ 23}$,
J.\thinspace Allison$^{ 16}$,
K.J.\thinspace Anderson$^{  9}$,
S.\thinspace Anderson$^{ 12}$,
S.\thinspace Arcelli$^{ 17}$,
S.\thinspace Asai$^{ 24}$,
S.F.\thinspace Ashby$^{  1}$,
D.\thinspace Axen$^{ 29}$
% ...
}\end{center}\bigskip
\bigskip
$^{  1}$School of Physics and Astronomy, University of Birmingham,
Birmingham B15 2TT, UK
\newline
$^{  2}$Dipartimento di Fisica dell' Universit\`a di Bologna and INFN,
I-40126 Bologna, Italy
\newline
% ...
$^{  8}$CERN, European Organisation for Particle Physics,
CH-1211 Geneva 23, Switzerland
\newline
$^{  9}$Enrico Fermi Institute and Department of Physics,
University of Chicago, Chicago IL 60637, USA
\newline
% ...
$^{ 12}$Indiana University, Department of Physics,
Swain Hall West 117, Bloomington IN 47405, USA
\newline
% ...
$^{ 16}$Department of Physics, Schuster Laboratory, The University,
Manchester M13 9PL, UK
\newline
$^{ 17}$Department of Physics, University of Maryland,
College Park, MD 20742, USA
\newline
% ...
$^{ 23}$Department of Physics and Astronomy, Tel Aviv University,
Tel Aviv 69978, Israel
\newline
$^{ 24}$International Centre for Elementary Particle Physics and
Department of Physics, University of Tokyo, Tokyo 113-0033, and
Kobe University, Kobe 657-8501, Japan
\newline
% ...
$^{ 29}$University of British Columbia, Department of Physics,
Vancouver BC V6T 1Z1, Canada
\newline
% ...
 Debrecen, Hungary.
\newline

\newpage
\section{Introduction}

The OPAL detector at LEP has collected more than 185~\pb\ 
of \epm\ collision data at centre-of-mass energies in the vicinity 
of $189$~GeV. These data are used to search for 
neutral Higgs bosons~\cite{higgs} 
within the framework of the Standard Model (SM)~\cite{sm}, 
and within the Minimal Supersymmetric extension of the 
Standard Model (MSSM)~\cite{mssm}.
Searches have been performed for 
the ``Higgs-strahlung'' process, 
\ee\ra\ho\Zo\ra\ho\ff, where \ho\  is either the SM 
Higgs boson \Hosm\ or the lightest neutral CP-even Higgs boson
in the MSSM, and
\ff\ is a fermion-antifermion pair from \Zo\ decay.
For the \ho\nunu\ (\ho\ee) final state, the contribution from 
the WW (ZZ) fusion process, which is only a small part of the total
production except near the kinematic threshold of the \ee\ra\ho\Zo\ process,
is also taken into account.
We have also searched for the MSSM process
\ee\ra\Ao\ho, where \Ao\ is the
CP-odd Higgs boson.

In this paper, only the dominant decays of the
neutral Higgs bosons (\ho\  and \Ao)
into \bb\  pairs and \tautau\  pairs are considered. 
We also consider the MSSM decay \ho\ra\Ao\Ao\ with \Ao\ra\bb .
Searches for Higgs boson decays into
SUSY particles are not presented in this paper.
OPAL searches at centre-of-mass energies up to 183 GeV
have been reported in~\cite{pr183,smpaper172,mssmpaper172}
for the neutral Higgs bosons of the SM and the MSSM.
For the SM Higgs boson a lower mass bound of $\mh>88.3$~GeV was obtained
at the 95\% confidence level from the lower-energy searches.
The searches presented here are similar in procedure 
to our previous searches at $\sqrt{s}\approx$183~GeV~\cite{pr183}; in this paper only 
significant changes from our previous analyses are described in detail.
In particular, we have significantly improved our b-tagging algorithm;
the new algorithm is described in detail in Section~\ref{sect:btag}.  

We have optimised the selection requirements in each of our searches
separately to maximise the sensitivity to Higgs boson production.  In each
case, the mass limit we would expect to set in the absence of a Higgs
signal, computed using Monte Carlo simulations of signal and background,
has been maximised.  When the distribution of a variable in the signal Monte
Carlo is needed as an input to a selection technique, we use a mixture of
Higgs masses near the expected limit in order to optimise
the analyses over the range of Higgs masses under study.

Recent searches for neutral Higgs bosons performed by the other LEP 
collaborations are listed in~\cite{higgs_otherlep}. The combined mass limit
by the four LEP collaborations
for the SM Higgs boson and a combination of the MSSM Higgs boson searches 
using data taken at $\sqrts \le 183$~GeV are reported in~\cite{lep_higgs}.

\section{OPAL Detector, Data Sets and Monte Carlo Samples}\label{sect:detector}
The present analysis is based on data collected with the OPAL 
detector~\cite{detector} during 1998 at a luminosity-weighted average
centre-of-mass energy of 188.6~GeV.  The searches presented here
use subsets of the data sample for which the necessary
detector components were fully operational.  Approximately 170~\pb\ were
analysed, varying $\pm$2\% from channel to channel, depending on
the subdetectors required.

The OPAL detector has nearly complete solid angle coverage.
The central tracking detector consists of a high-resolution
silicon microstrip vertex detector~\cite{simvtx} which immediately
surrounds the beam pipe.  Its coverage in
polar angle\footnote{
OPAL uses a right-handed
coordinate system where the $+z$ direction is along the electron beam and
where $+x$ points to the centre of the LEP ring.  
The polar angle $\theta$ is
defined with respect to the $+z$ direction and the azimuthal angle $\phi$
with respect to the $+x$ direction.}
is $|\cos\theta|<0.9$.   The silicon microvertex detector
is followed by a high-precision 
vertex drift chamber,
a large-volume jet chamber, and $z$--chambers to measure the $z$ coordinate
of tracks, all in a uniform
0.435~T axial magnetic field. The lead-glass electromagnetic calorimeter
with a presampler is located outside the magnet coil.  It provides,
in combination with the forward calorimeters,
the forward scintillating tile counter (the ``MIP plug'')~\cite{llpaper},
and the silicon-tungsten luminometer~\cite{sw}, geometrical acceptance
down to 25~mrad from the beam direction.  The silicon-tungsten luminometer
serves to measure the integrated luminosity using small-angle Bhabha
scattering events~\cite{lumino}.
The magnet return yoke is instrumented with streamer tubes and thin gap
chambers for hadron calorimetry; it is surrounded by several layers 
of muon chambers.

Events are reconstructed from charged-particle tracks and
energy deposits (``clusters") in the electromagnetic and hadron calorimeters.
The tracks and clusters must pass a set of quality requirements
similar to those used in~\cite{higgsold}.
In calculating the total visible energies and momenta, $E_{\mathrm vis}$
and $\vec{P}_{\mathrm vis}$, of events and
individual jets, ``energy-flow objects'' are formed from the tracks
and calorimeter clusters\cite{lep2neutralino}. 
A track and its associated clusters are grouped
into a single energy-flow object after correcting the clusters for the
estimated energy deposited by the particle leaving the track.
Unassociated clusters are left as energy-flow objects.  This procedure
prevents double-counting of the energies of particles which leave signals in
more than one subdetector.

A variety of Monte Carlo samples has been generated in order to estimate the
detection efficiencies for Higgs boson production and SM background processes.
Higgs production is modelled with the HZHA generator~\cite{hzha}
for a wide range of Higgs masses.  The size of these samples
varies from 500 to 10,000 events.
The background processes are simulated, with typically more than 50 times the
statistics of the collected data,
by the following event generators:
PYTHIA~\cite{pythia} (\Zgs\ra\qq($\gamma$)), 
grc4f~\cite{grc4f} and EXCALIBUR~\cite{excalibur} (four-fermion processes 
(4f)),
BHWIDE~\cite{bhwide} (\ee$(\gamma)$),
KORALZ~\cite{koralz} (\mm$(\gamma)$ and \tautau$(\gamma)$),
and PHOJET~\cite{phojet}, HERWIG~\cite{herwig}, and
Vermaseren~\cite{vermaseren} (hadronic and leptonic two-photon processes
($\gamma\gamma$)).
The hadronisation process is simulated with 
JETSET~\cite{pythia} with parameters
described in~\cite{opaltune}.
The cluster fragmentation model in HERWIG is used
to study the uncertainties due to fragmentation and hadronisation.
For each Monte Carlo sample, the detector response to the generated
particles is simulated in full detail~\cite{gopal}.

\section{Improved b-Tagging}\label{sect:btag}

  The predominant decay mode of the
  SM Higgs boson in the mass range under study is expected
  to be to pairs of b quarks; the same is true of the neutral Higgs
  bosons of the MSSM in large areas of the model's parameter space.
  Therefore, efficient and pure tagging of b quarks 
  is an essential technique in searches for these particles.
  We have developed a tagging method~\cite{pr183} that
  uses the three nearly independent
  techniques of lifetime, high-$p_t$ lepton and kinematic tagging
  to identify jets containing b hadron decays and to minimise the
  contamination from jets which do not.
  Artificial Neural Networks (ANN's) have been introduced
  to combine optimally several lifetime-sensitive tagging variables and
  also to combine kinematic variables in the jet-kinematics
  part of the tag.  For each jet, the outputs of the lifetime
  ANN, the kinematic ANN,
  and the lepton tag are combined into a likelihood variable ${\cal B}$ 
  which discriminates b-flavoured jets from c-flavoured and uds-flavoured jets.
  While the basic scheme of the tagging procedure has been kept unchanged
  with respect to our previous procedure~\cite{pr183}, a significant improvement
  in the performance has been achieved for this analysis
  by introducing input variables with greater sensitivity.

  In order to improve the performance of the lifetime-sensitive part of the tag,
  we have developed a new algorithm to identify displaced vertices.
  An ANN applied to individual tracks (track-ANN) has been trained, 
  using Z$^0$-pole Monte Carlo samples, 
  to discriminate between tracks from the primary vertex and those
  from secondary vertices.  It uses, among other inputs, the
  impact parameter of the track with respect to the primary vertex and 
  the transverse momentum with respect to the jet axis.
  The tracks belonging to a given jet are ranked in descending order of
  their probability of having come from a secondary vertex, according 
  to the track-ANN output.  Using the first six tracks (or all tracks
  in the jet if there are fewer than six), a ``seed''
  vertex is formed using the technique described in~\cite{btag1}.  This
  technique first forms a vertex from the input tracks, and then removes the
  track which contributes the most to the vertex $\chi^2$.  This procedure
  is repeated until no track contributes more than 5 to the $\chi^2$, and the
  seed vertex is the common fit origin of the tracks.  The remaining
  tracks in the jet are then tested to see if they may be added to the seed
  vertex, based on their contributions to the vertex $\chi^2$.

  In addition to identifying displaced vertices, we introduce a new
  method for combining track impact
  parameters in order to gain b-tagging efficiency in events where the 
  secondary vertices are less distinct.  In this method,
  the impact parameter significances $S^{r\phi}$ and $S^{rz}$,
  in the $r\phi$ and $rz$ projections, respectively, are formed by dividing
  the track impact parameters by their estimated errors.
  The distributions of $S^{r\phi}$
  and $S^{rz}$ for each quark flavour from Z$^0$-pole Monte Carlo 
  are used as the probability density functions (PDF's)
  $f^{r\phi}_{q}$ and $f^{rz}_{q}$ ($q$=uds, c and b).
  The combined estimator $F_{q}$ for each quark flavour $q$ is computed
  by multiplying the  $f^{r\phi}_{q}$ and $f^{rz}_{q}$ for all tracks
  passing the quality requirements.
  The final estimator ${\cal L}_{\mathrm IP}$ is obtained as the
  ratio of $F_b$ and the sum of $F_{uds}$, $F_c$ and $F_b$. 

 The following four variables are used as inputs to the lifetime
 ANN.  They replace the five variables used in our 183 GeV analysis~\cite{pr183}.
\begin{itemize}
\item The vertex significance likelihood ($\cal{L}_{\mathrm V}$):
  The likelihood for the vertex significance is computed analogously to
  the ${\cal L}_{\mathrm IP}$ above, using the decay length significance
  of the secondary vertex rather than the impact parameter significance
  of the tracks.  Because the decay length significance of a secondary vertex
  depends strongly on the number of tracks in it, the PDF's for each flavour
  are computed separately for each value of the charged multiplicity of
  the secondary vertex.
\item The reduced significance likelihood ($\cal{R}_{\mathrm V}$):
  To reduce sensitivity to single mismeasured tracks,
  the track with the largest impact parameter significance with respect
  to the primary vertex is removed from
  the secondary vertex candidate and the remaining tracks are used to
  recompute the likelihood $\cal{L}_{\mathrm V}$.
  If the original vertex has only two tracks, the function is
  calculated from the impact parameter significance of the remaining track.
\item The combined impact parameter likelihood ($\cal{L}_{\mathrm IP}$) described above.
\item The reduced impact parameter likelihood ($\cal{R}_{\mathrm IP}$):
   The track having the largest impact parameter significance has been
   removed in the calculation of $\cal{L}_{\mathrm IP}$.
\end{itemize}

In the jet-kinematics part of the tag,
our previous procedure used only the boosted sphericity of the jet as an input
to the final likelihood.  This variable has been replaced with
three new variables, which are combined with a separate ANN.
These three inputs are
1) the number of energy-flow objects around the central 
part of the jet,
2) the angle between the jet axis and its boosted sphericity axis, and
3) the $C$-parameter~\cite{cpar} for the jet boosted back to its rest frame.
The high-$p_t$ lepton tag has not been changed.

\begin{figure}[p]
\vspace*{-2cm}
\centerline{\epsfig{file=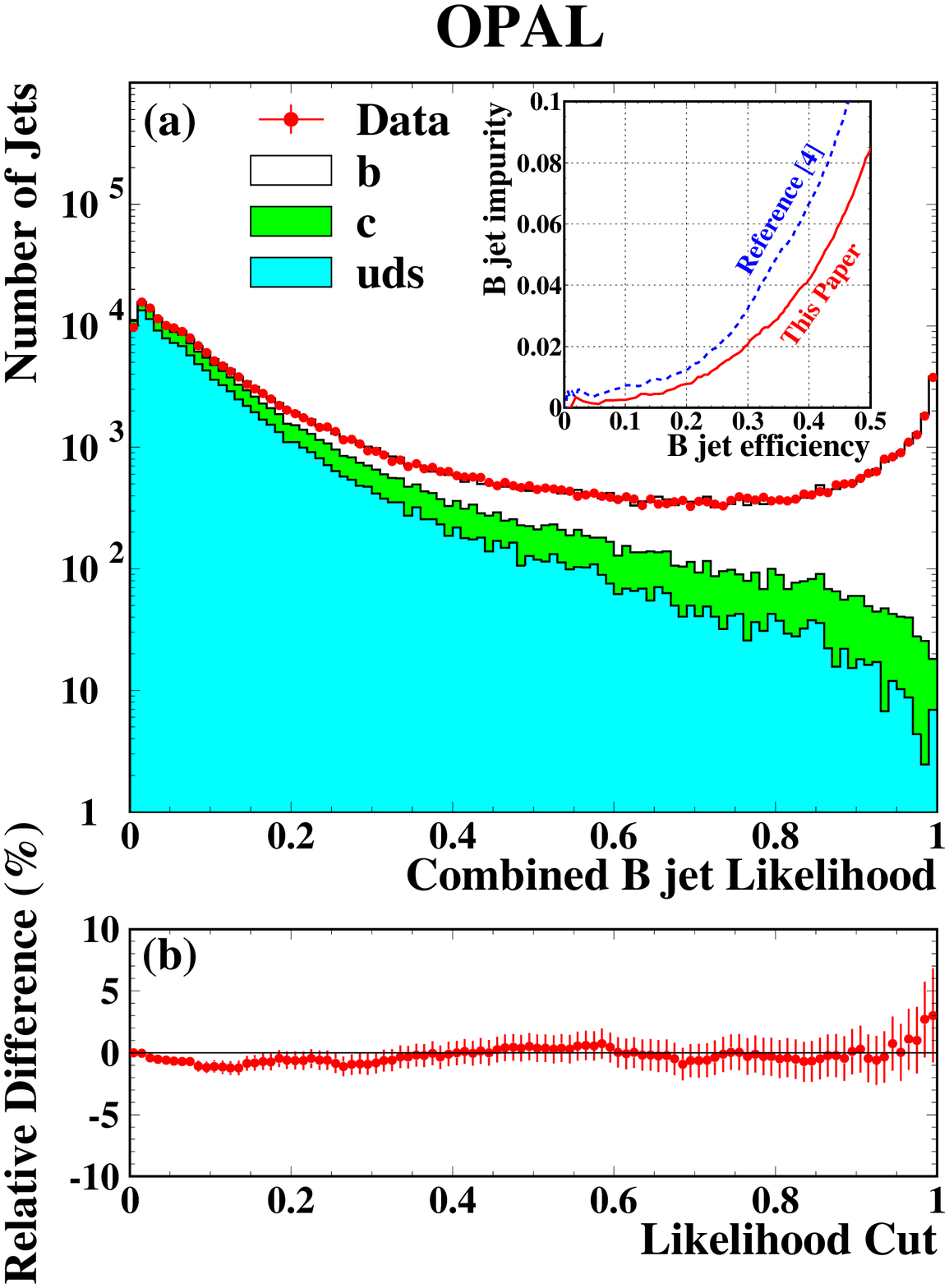,width=0.95\textwidth}}
\caption[]{\label{fig:btag}\sl
B-tagging performance and modelling.
(a): the distribution of the output of 
the b-tagging algorithm, $\cal B$, for jets in
calibration data taken at $\sqrt{s}=m_{\mathrm{Z}^{0}}$, 
compared to the Monte Carlo expectation. 
The data distribution is given by the points, with error bars smaller
than the plot symbols.
The open histogram shows the distribution of $\cal B$ for b-flavoured jets,
and the dark (light) grey histogram shows the contribution from 
c (uds) flavoured jets, expected in a Monte Carlo simulation.
Inset: The b-tagging performance in the present work for hadronic jets in
\Zo\ decay compared with our previous version~\cite{pr183}.
(b): the relative difference of the tagging rates for jets opposite
to b-tagged jets in \Zo\ calibration data and Monte Carlo as a function of the likelihood cut.
The points are statistically correlated because the tagging rate accumulates
events on one side of the likelihood cut.  
}
\end{figure}

The outputs from the lifetime ANN, the
jet-kinematics ANN and the high-$p_t$ lepton tag
are combined with an unbinned likelihood calculation as described in~\cite{pr183}, and
the final output ${\cal B}$ is computed for each jet.
Figure~\ref{fig:btag}(a) shows the distribution of
${\cal B}$ for calibration data 
taken at $\sqrt{s}=m_{\mathrm{Z}^{0}}$ in 1998; it also
shows the improved performance 
of the new b-tagging algorithm,
compared with the previous version~\cite{pr183}.
For the same efficiency, the new b-tag has about two-thirds of the background
acceptance of the previous tag.
The tagging efficiency for b-flavoured jets
has been verified to an accuracy of 1\% with a
double-tagging technique using the \Zo -calibration data collected
with the same detector configuration and operating 
conditions as the high-energy data; a comparison of the
results of this double-tagging test between the data and Monte Carlo is
shown in Figure~\ref{fig:btag}(b).
Using the high energy data, the expectation from the SM Monte Carlo agrees
very well, within a relative statistical uncertainty of 5\%,
for samples enriched in \Zgs$\rightarrow{\mathrm q\bar{q}}$ processes with and without hard
initial-state photon radiation. 
For the lighter flavours, the efficiency has been examined by vetoing b~flavour
in the opposite hemisphere, and the resulting estimate of the background
is found to be described by Monte Carlo within an accuracy of
 5--10\%.  This constitutes
one component of the systematic uncertainty in the background estimates
of the search
channels presented here.
The efficiency for tagging lighter flavours has also
been checked by computing ${\cal B}$ for a high-purity sample of light-flavour jets
in ${\mathrm W^{+}W^{-}\rightarrow q{\overline q}}\ell{\overline \nu}$ decays;
there is no evidence of mismodelling of the light-flavour tag rate within the
statistical precision of the test.

\section{Searches for ${\protect \boldmath \ee\ra\ho\Zo}$}
\label{sect:zhsearches}

Throughout this section, \ho\ denotes both the Standard Model Higgs boson
\Hosm\ and the lightest CP-even Higgs boson in the MSSM.
We have searched for the process \ee\ra\ho\Zo\ in the following final
states: \ho\Zo\ra\bb\qq\  (the four-jet channel), \ho\Zo\ra\bb\nn\ 
(the missing-energy channel),
\ho\Zo\ra\bb\tautau\ and \tautau\qq\
(the tau channels), \ho\Zo\ra\bb\ee\ and \bb\mm\ (the electron and
muon channels).
The selections for all channels are similar to those described in~\cite{pr183};
their main features and differences from our previous publication
are described in Sections~\ref{sect:sm4jet} through~\ref{sect:smlept}.
These selections are also sensitive to the \ho\Zo\ra\Ao\Ao\Zo\  processes of the
MSSM, and the procedure for incorporating this additional information into our
limits is described in Section~\ref{sect:hzaa}.


\subsection{\label{sect:sm4jet} The Four-Jet Channel}
The selection for the four-jet channel has been modified with respect to
our previous publication in order to improve the
sensitivity to Higgs signals at higher masses.
In addition to using the improved b-tagging algorithm described above, 
we have improved the grouping of final-state particles into four jets 
and the assignment of pairs of these jets to the \Zo\ and \ho\ bosons.

The correct assignment of particles to jets plays an essential role
in reducing one of the main backgrounds, \WW\ra\qq\qq, and also
in accurately reconstructing the Higgs boson mass.
The Durham algorithm~\cite{durham} 
is used to find four jets in each event, {\it i.e.},
the resolution\ parameter $y_{\mathrm cut}$ is chosen to be between
$y_{34}$ and $y_{45}$, where $y_{34}$ is the transition point
from three to four jets, and 
$y_{45}$ is the transition point from four to five jets.
These jets are used as reference jets in the following 
procedure.
Each particle is reassociated to the jet 
having the smallest ``distance'' in the E0 scheme~\cite{OPAL-E0} of 
the JADE algorithm~\cite{JADE-jet} to the particle.  This ``distance'' is
defined to be
$E^{i}_{jet} \cdot E_{particle} \cdot (1-\cos{\theta_{i}})$, 
where $E^{i}_{jet}$ is the energy of the $i^{\mathrm th}$
 reference jet ($i$=1 to 4), 
$E_{particle}$ is the energy of the particle (charged track and/or calorimeter cluster,
corrected for double-counting) 
and $\theta_{i}$ is the angle between the $i^{\mathrm th}$ reference jet 
and the particle. This reassociation procedure reduces the fraction of
wrong association of particles to jets, which improves by 
about 10\% the di-jet mass resolution for the \WW\  background as well as for
Monte Carlo Higgs signals, before kinematic fitting.
The rearranged jets are used in cuts~5 and~6 of the preselection and
the final likelihood selection described below.  This procedure is also
used for the four-jet \ho\Ao\ channel described in Section~\ref{sect:habbbb}.

The preselection is exactly the same as that used in~\cite{pr183}; it
is designed to retain
only four-jet-like events.  The requirements are:
 
\begin{enumerate}
\item  Events must satisfy the hadronic final state requirement of
  Reference~\cite{l2mh}.
\item The effective centre-of-mass energy $\sqrt{s^\prime}$~\cite{l2mh} 
      must be at least 150 GeV.
\item The value of $y_{34}$ in the Durham algorithm must exceed 0.003.
\item The event shape $C$--parameter~\cite{cpar} must be larger than 0.25.
\item Each of the four jets must have at least two tracks.
\item The $\chi^2$ probabilities must be larger than 10$^{-5}$ both for 
 a four-constraint (4C) kinematic fit, which requires energy and momentum 
conservation,
 and a five-constraint (5C) kinematic fit, additionally constraining 
the invariant mass of
one pair of jets to $m_{{\mathrm Z}^0}$~\cite{smpaper172}.
The algorithm used to select the jet pair which is constrained
to the \Zo\ mass is described below.
\end{enumerate}

There are six possible ways to assign four jets in pairs to the \Zo\ and the \ho.
An algorithm based on a likelihood technique 
has been developed in order to select the most Higgs-like combination.
This algorithm makes use of b-tagging
variables and the probability $p_{\mathrm 5C}$
that the 5C kinematic fit would have had the observed $\chi^2$ or larger.
For each of the six jet-assignment combinations,
we form two variables from the b-tagging output as follows:
(1) ${\cal B}_1 \cdot {\cal B}_2$ for jets
assigned to the Higgs boson, and
(2) $(1-{\cal B}_3) \cdot (1-{\cal B}_4) $ for jets
assigned to the \Zo.
These two variables, together with log($p_{\mathrm 5C}$) for the jet combination,
are used as inputs to the likelihood calculation, which is designed to
distinguish between the three cases of correct jet assignment, swapped pairings in which
the two jets from the \ho\ are wrongly assigned to the \Zo,
and other combinations, for which one jet from the \ho\ is assigned to the \Zo.
For each of the possible assignments of pairs of jets to the \Zo\ and the \ho\
 with $p_{\mathrm 5C} > 10^{-5}$,
three likelihoods, ${\cal L}^{correct}$, ${\cal L}^{swapped}$ and  ${\cal L}^{others}$, 
are calculated using the PDF's of the input variables in signal Monte Carlo
events
passing the preselection.  The relative likelihood 
$ {\cal L} = 
{\cal L}^{correct}/\left({\cal L}^{correct}+{\cal L}^{swapped}+{\cal L}^{others}\right)$
is then formed for each combination and 
the combination yielding the largest value of $\cal{L}$ is chosen. The
reconstructed Higgs mass is taken to be the mass obtained
from the 5C kinematic fit for this combination.

With this algorithm, the rate of correct jet pairings for selected signal
events with $m_h = 95$ GeV is 71\%, compared to the rate of 36\% obtained 
with the method used in our 183 GeV analysis, which uses only
$p_{\mathrm 5C}$ for the assignment.
This improvement in the correct jet-pairing reduces the
fraction of combinations for which the reconstructed \ho\ and \Zo\ masses
have been swapped.  The fraction of signal Monte Carlo events with $m_h = 95$~GeV
for which the reconstructed Higgs mass is in the Gaussian core centered on 95~GeV
increases by 10\% when
this improved jet pairing is used, as compared to the previous 
pairing.
The correct jet pairing provides 
better assignment of the b-tag variables to the Higgs candidate and improves the separation 
of the \ho\Zo\ signal from \Zo\Zo\ background.

\begin{figure}[p]
\centerline{\epsfig{file=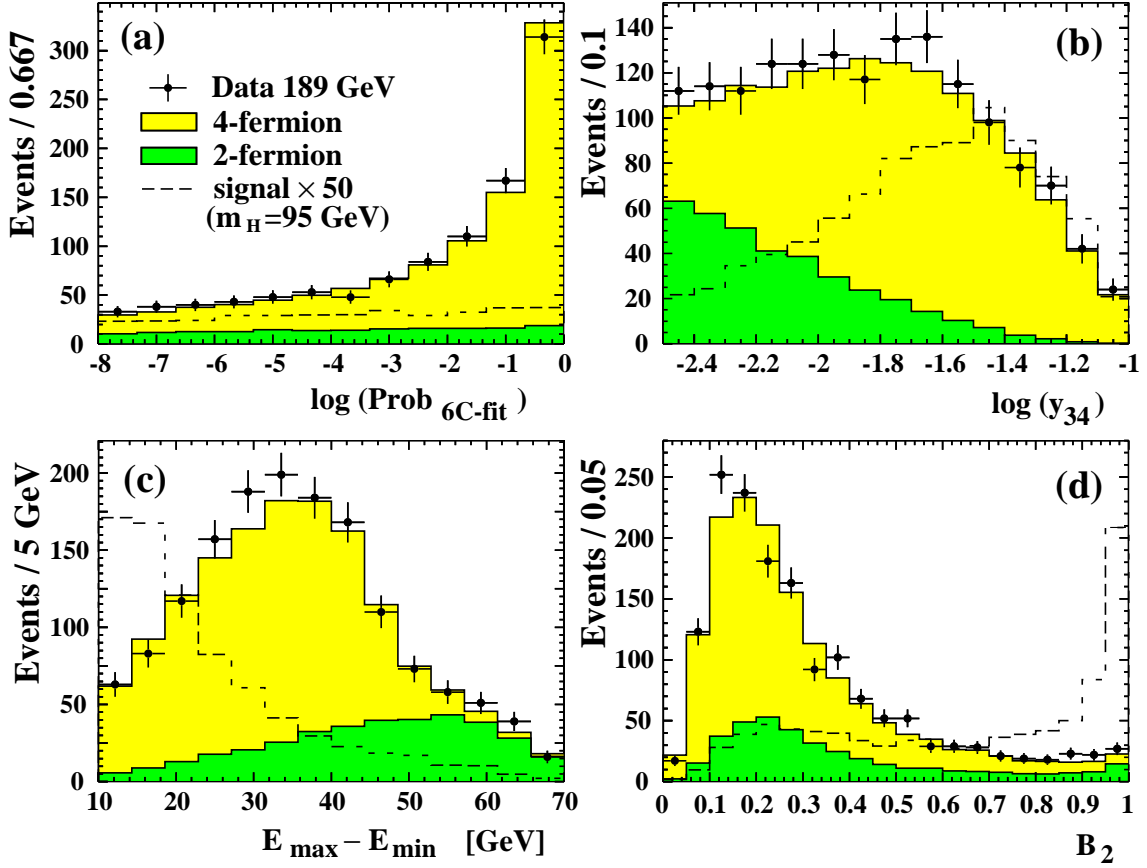,width=0.95\textwidth}}
\caption[]{\label{fig:zhqbinputs}\sl
Input variables for the four-jet channel likelihood selection:
(a) the fit probability of a 6C kinematic fit which requires energy and momentum
conservation and both dijet masses to be equal to $m_{\mathrm{W}}$;
(b) the logarithm of the jet resolution parameter $y_{34}$;
(c) the difference between the energies of the jets with the highest
and lowest energies;
(d) the second largest jet b-tag.
OPAL data are
indicated by points with error bars, four-fermion backgrounds
by the light grey histograms, and two-fermion backgrounds by the
dark grey histograms.  The estimated contribution from a 95~GeV Higgs is 
shown with dashed histograms; it has been scaled by a factor of 50.}
\end{figure}
\begin{figure}[p]
\centerline{\epsfig{file=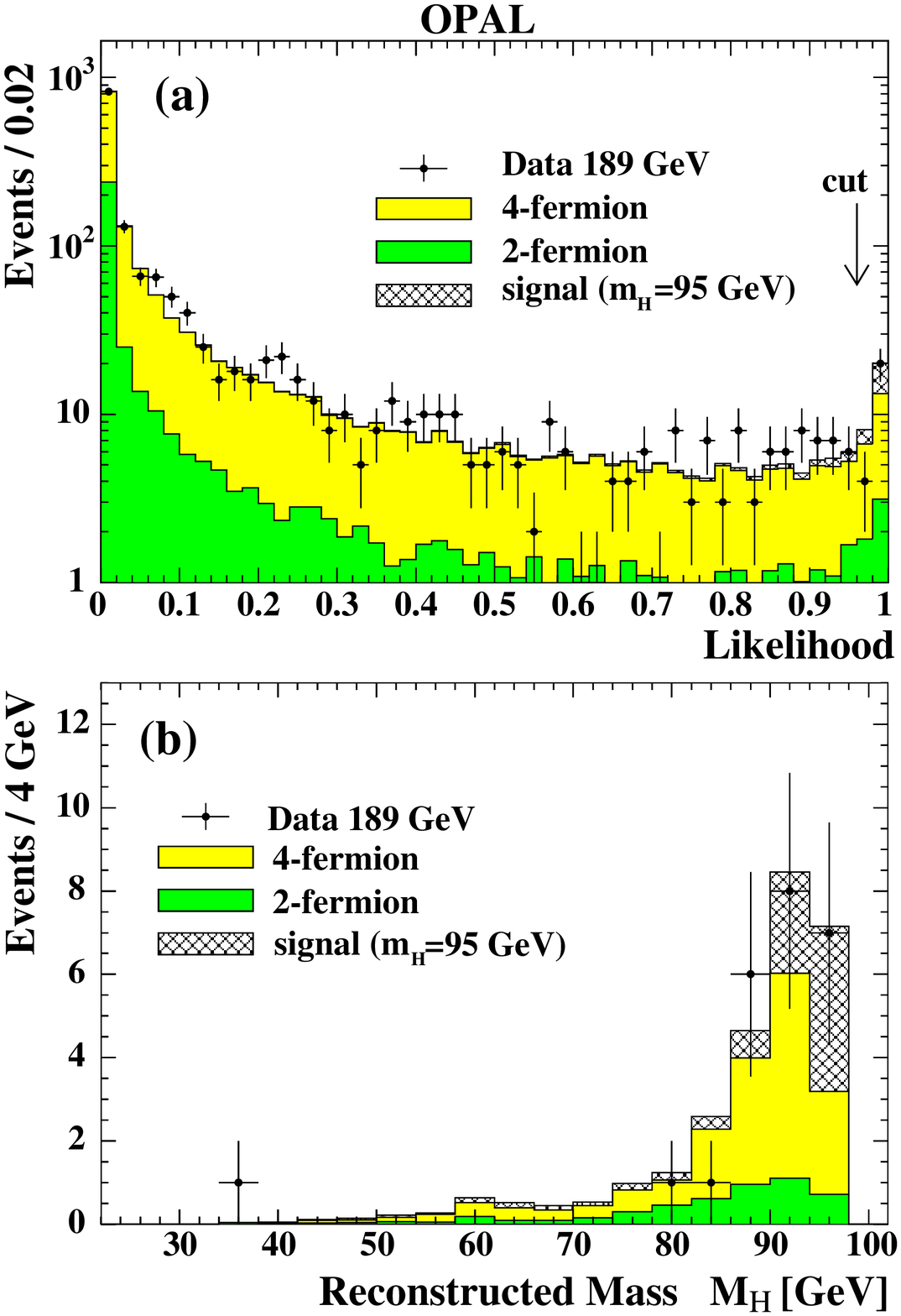, width=0.95\textwidth}}
\caption[]{\label{fig:zhqblike}\sl
Four-jet channel: the likelihood distribution (a) and 
the reconstructed candidate masses (b).
OPAL data are
indicated by points with error bars, four-fermion backgrounds
by the light grey histograms, and two-fermion backgrounds by the
dark grey histograms.  Also shown is the contribution expected from
a 95~GeV Higgs boson (hatched histograms).}
\end{figure}

After the preselection, the eight variables described in~\cite{pr183}
are combined using a likelihood method~\cite{smpaper172}.
The distributions of four of these variables are shown in 
Figure~\ref{fig:zhqbinputs} for the
OPAL data and corresponding SM background simulations.
The likelihood distribution ${\cal L}^{\mathrm HZ}$
and the distribution of the reconstructed Higgs boson candidate mass
are shown in Figure~\ref{fig:zhqblike}.  Higgs boson candidate events
are required to have ${\cal L}^{\mathrm HZ}>0.96$.
The numbers of observed and expected events after each selection 
step are given in Table~\ref{tab:smflow}.

After the likelihood selection, 24 candidate events 
are retained.  The expected SM background is 
19.9$\pm$0.8(stat.)$\pm$2.9(syst.) events, computed using
Monte Carlo simulations. Within the contribution from 
\Zo\Zo\ra\qq\qq\ (10.4 events), more than 90\% contain
at least one \Zo\ which decays into \bb. 
The efficiency for $\mh = \mbox{95 \Gc\ }$ is
(47.0$\pm$0.8(stat.)$\pm$1.6(syst.))\%.
The systematic errors for the signal selection efficiencies and 
the background estimates are assigned in the same way as 
described in~\cite{pr183}. The signal selection efficiency as a 
function of the \ho\ mass is given in Table~\ref{tab:smsummary}, which summarises
the performance of all the SM channels.

\begin{figure}[p]
\centerline{\epsfig{file=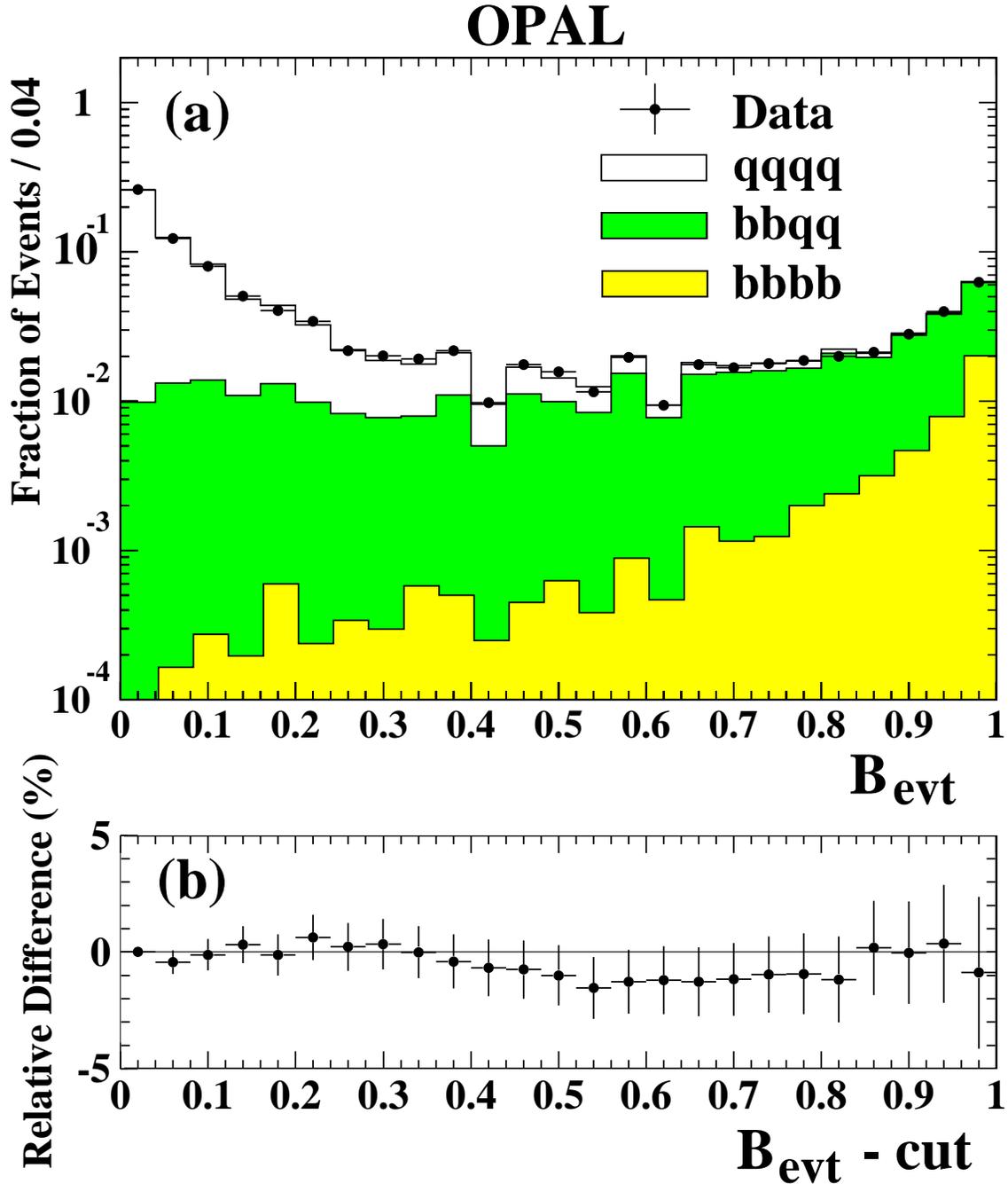,width=0.95\textwidth}}
\caption[]{\label{fig:zhqbover}\sl
Four-jet channel:
(a) the $B_{\mathrm{evt}}$ distribution for 
overlaid pairs of hadronic \Zo\ decay events 
(described fully in the text). 
OPAL data are indicated by points (the error bars are smaller
than the points), and the Monte Carlo simulation
by the histograms with the contributions from final states 
containing four b-quarks (light), two b-quarks (dark) and only
light flavoured quarks (open).
The discontinuous behaviour in the central bins is due to the discrete
binning of the two b-tag variables used as inputs to the likelihood.
(b) The relative difference between OPAL data and the Monte Carlo simulation
for the event b-tagging rate as a function of the cut on 
$B_{\mathrm{evt}}$ for pairs of overlaid \Zo\ decays.}
\end{figure}

In order to verify the correct modelling of the b-tag algorithm in the 
high-multiplicity four-jet environment, the following cross-check has been
performed. The measured charged particle tracks and calorimeter clusters
of pairs of hadronic \Zo\ decay events recorded in 1998 at $\sqrts = \mZ$ have been 
overlaid to form pseudo-events which have topologies very similar to
$\ee\ra\Zo\Zo\ra\qq\qq$ events at LEP2 energies. In order to model the
b~tag properly, the charged track parameters of the second
\Zo\ event have been adjusted such that the reconstructed primary vertices
of both overlaid events match. The same procedure has been applied to
simulated hadronic \Zo\ decays. These pseudo-events have been passed through the
same analysis chain as the high energy data and Monte Carlo. The same
preselection as described above has been applied.  
The event b-tag likelihood $B_{\mathrm evt}$
is calculated using as inputs 
the two largest jet-wise b-tag variables -- the same b-tag
variables used in the \ho\Zo\ selection.
The distribution for $B_{\mathrm evt}$
is shown in Figure~\ref{fig:zhqbover} for overlaid data and overlaid Monte Carlo
as well as their relative difference in the event tagging rate as a function
of the cut on $B_{\mathrm evt}$.  The data and the Monte Carlo model agree to within
2\%, independent of the cut on $B_{\mathrm evt}$.  Hence, no additional
systematic error has been assigned on the signal efficiency as a result 
of this cross-check.

\subsection{The Missing-Energy Channel}
\label{sect:sm-miss}
The preselection of the missing-energy analysis is designed to enhance
a signal characterised by two hadronic jets and missing
energy in the presence of
radiative \Zo\  events, untagged
two-photon events, and \WW\ra\qq\lnu\  events. 
These Standard Model backgrounds also can produce final states with
two hadronic jets and missing energy, but they can be rejected because
their missing momentum is predominantly along the beam axis (in the case
of the radiative \Zo\ events and untagged two-photon events), or there
is a high-momentum lepton (in the case of \wwln\ events).
The preselection is described below.

The initial requirements are intended to guarantee well measured events
with a reduced two-photon and $\mathrm{Z}/\gamma^*$ background: (1)
the number of tracks satisfying the quality requirements used in~\cite{higgsold}
must be greater than six and more than
20\% of all tracks;
there must be no significant energy in the forward detectors
as described in~\cite{smpaper172}; no hits in the MIP plug detector with
a significant charge deposition;
the total transverse momentum $P^t_{\mathrm{vis}}$ and the visible mass
$m_{\mathrm{vis}}$ must satisfy 
$5 \cdot P^t_{\mathrm{vis}} + m_{\mathrm{vis}} > 100$~GeV;
the total visible energy $E_{\mathrm{vis}} < 0.8\cdot \sqrts$;
and there must be less than 
50\% of the visible energy in the angular region $|\cos{\theta}| > 0.90$.
The following requirements reduce the $\mathrm{Z}/\gamma^*$ contribution:
(2) the polar angle of the missing momentum, $\theta_{\mathrm{miss}}$,
must satisfy $|\cos\theta_{\mathrm miss}|<0.95$ and
the $z$-component of the visible momentum must satisfy
$|P^z_{\mathrm vis}| < 35$~GeV;
(3) the tracks and clusters are grouped into two jets
using the Durham algorithm, and the directions of both jets 
are required to satisfy
$|\cos\theta_{\mathrm{jet}}|<0.95$;
(4) the acoplanarity angle\footnote{
The acoplanarity angle, $\phi_{\mathrm{acop}}$, of
two vectors is 180$^\circ$ minus the opening angle
between them in the plane transverse to the beam
direction.
} of the two jets, $\phi_{\mathrm{acop}}$, 
must be larger than $5^{\circ}$.
The following requirements reduce the contribution from four-fermion
processes:
(5) the missing mass $m_{\mathrm{miss}}$ must be consistent with \mZ:
$60~\mathrm{GeV}< m_{\mathrm{miss}} <120$~GeV; and
(6) no identified isolated leptons, as defined in~\cite{smpaper172}, may
appear in the event.

\begin{figure}[p]
\vspace*{-2cm}
\centerline{\epsfig{file=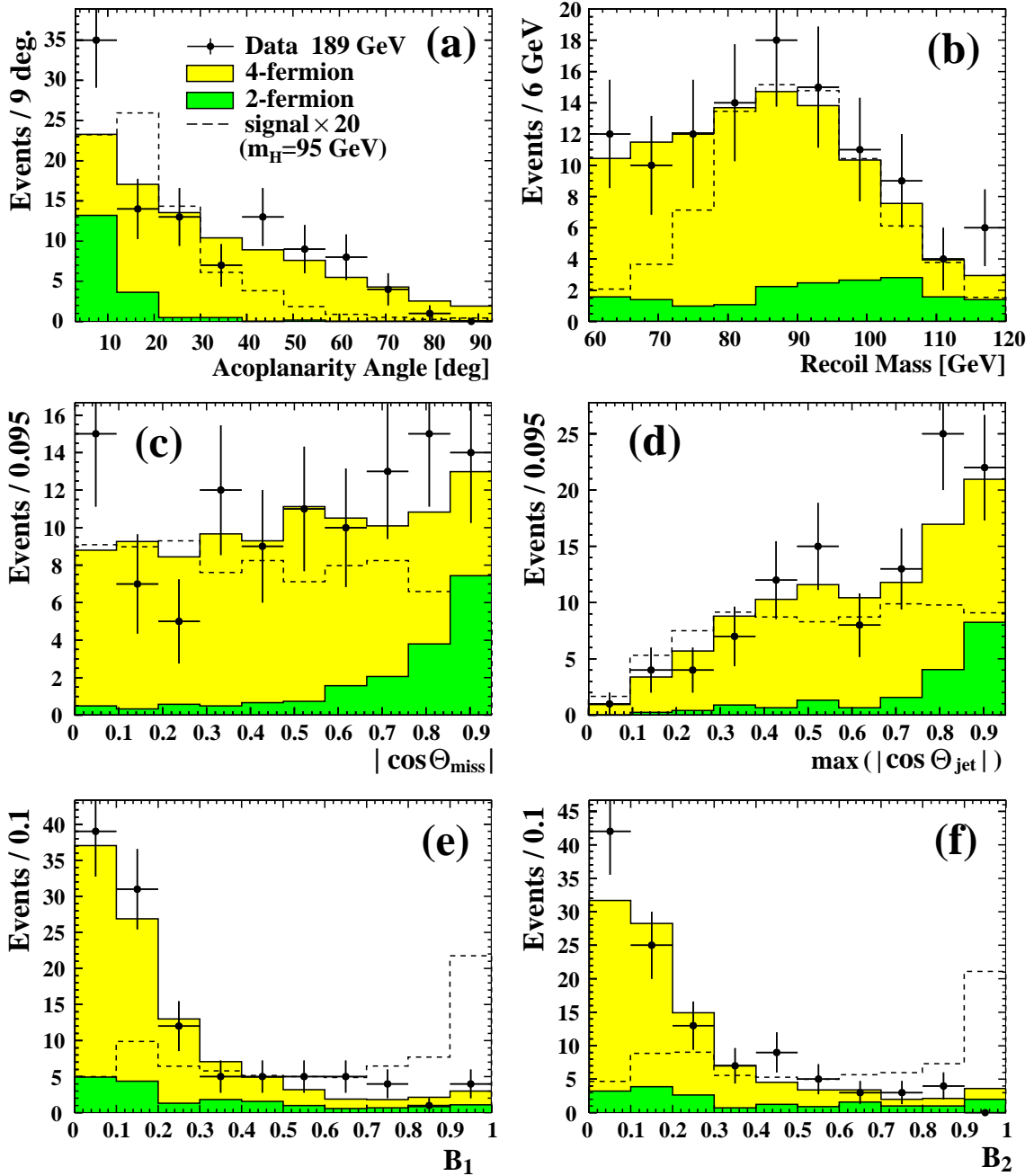,width=0.95\textwidth}}
\caption[]{\label{fig:missinputs}\sl
Input variables to the missing-energy channel likelihood selection:
(a) the acoplanarity angle;
(b) the invariant mass recoiling against the hadronic system;
(c) the cosine of the polar angle of the missing momentum;
(d) the larger cosine of the two jet polar angles;
(e) and (f) the b-tags of the two jets.
  OPAL data are
indicated by points with error bars, four-fermion backgrounds
by the light grey histograms, and two-fermion backgrounds by the
dark grey histograms.  The estimated contribution from a 95~GeV Higgs boson 
is shown with dashed histograms; it has
been scaled up by a factor of 20.}
\end{figure}

\begin{figure}[p]
\centerline{\epsfig{file=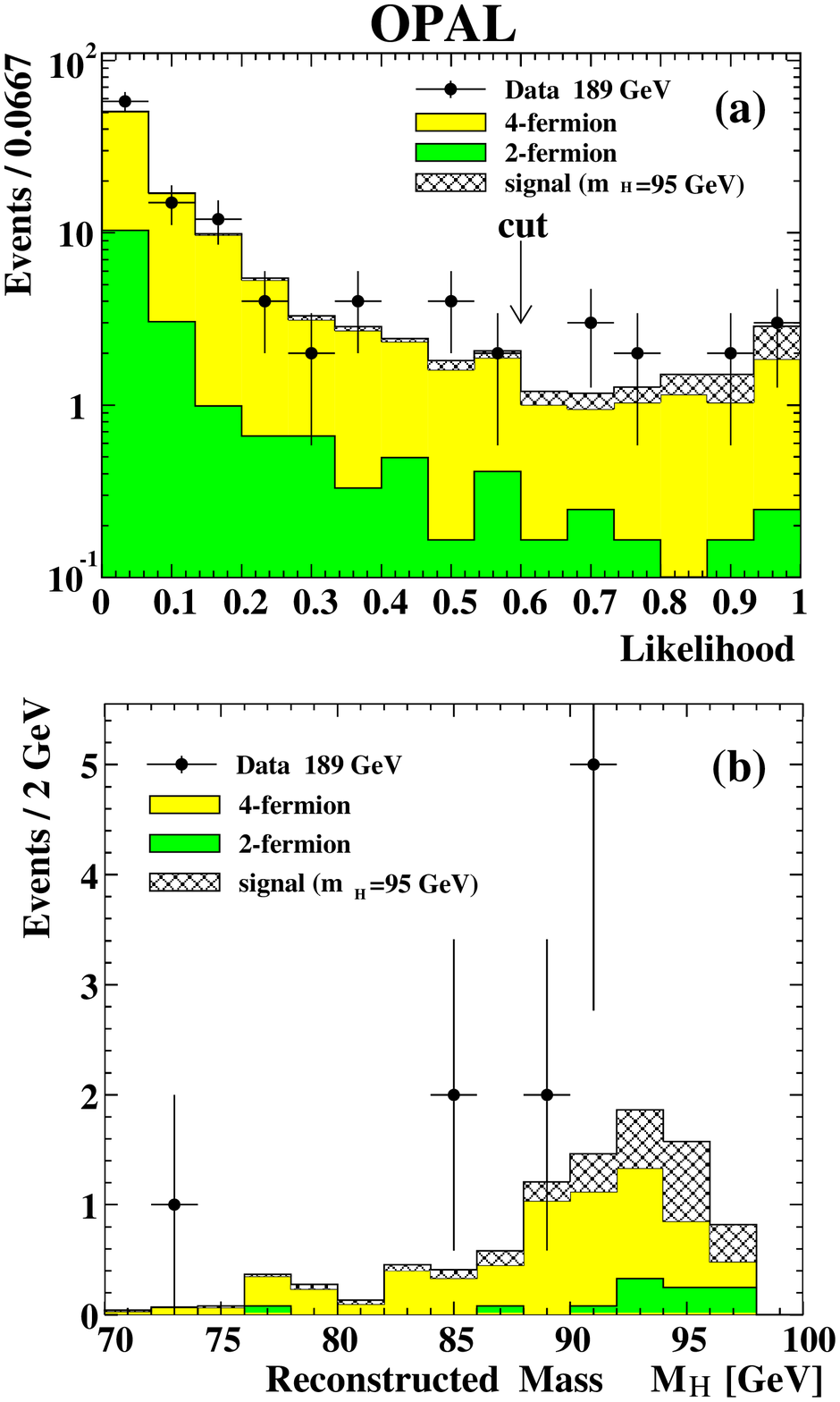, width=0.8\textwidth}}
\caption[]{\label{fig:misslike}\sl
Missing-energy channel: (a) the likelihood distribution and (b) 
the reconstructed candidate masses.
OPAL data are
indicated by points with error bars, four-fermion backgrounds
by the light grey histograms, and two-fermion backgrounds by the
dark grey histograms.  Also shown is the contribution expected from
a 95~GeV Higgs boson (hatched histograms).}
\end{figure}

The b-tag described in Section~\ref{sect:btag} is incorporated into the analysis
by combining ${\cal B}_1$ and ${\cal B}_2$, the b-tagging discriminants
of the two jets in the event, with four
kinematic variables, $|\cos\theta_{\mathrm{miss}}|$, 
${\mathrm max}({\cos\theta_{\mathrm jet}})$, $m_{\mathrm miss}$,
and $\phi_{\mathrm acop}$, using the same likelihood 
technique as used at lower energies~\cite{pr183}.
The high-$p_t$ lepton
tagging has been removed from the calculation of the jet b-tagging
discriminant variables \Dbi\ in order to avoid enhancing the
\wwln\ background.
Distributions of the kinematic and b-tag variables 
are shown in Figure~\ref{fig:missinputs}
for OPAL data and SM background simulations.
The signal likelihood is required to be larger than 0.60 for an event
to be selected as a Higgs boson candidate.
In Figure~\ref{fig:misslike}, the likelihood distribution and the
mass distribution for the selected 
candidate events are shown for the data,
SM backgrounds, and a simulated signal at $m_{\mathrm h}=95$~GeV.
The reconstructed Higgs mass
is evaluated using a kinematic fit constraining
the recoil mass to the \Zo\ mass. 

The numbers of observed and expected events after each selection 
step are given\footnote{
In the calculation of the efficiencies and backgrounds in the missing-energy
channel,
a 2.5\% relative reduction 
has been applied to the Monte Carlo estimates
in order to account for accidental vetoes 
due to accelerator-related backgrounds in the forward detectors.} 
in Table~\ref{tab:smflow}.
The selection efficiency estimated from the Monte Carlo for a 95 GeV Higgs
is (35.4$\pm$0.9(stat.)$\pm$0.9(syst.))\%. 
Ten events survive the selection, while
6.9$\pm$0.5(stat.)$\pm$0.6(syst.) events
are expected from SM background processes.
The systematic error evaluation is described in~\cite{pr183}.
The detection efficiencies as a function of the Higgs boson mass 
are listed in Table~\ref{tab:smsummary}.

%
\subsection{The Tau Channels}
\label{sect:smbbtt}
The tau channel selection consists of a preselection and 
tau lepton identification using an
ANN, the details of which are described in~\cite{pr183}.
The preselection requirements to select the signal with a
low contamination of $\mathrm{Z}/\gamma^*$ events are:
the event must be a high-multiplicity multihadronic event~\cite{l2mh},
$\left| \cos\theta_{\mathrm{miss}} \right| \leq 0.95$,
the missing momentum $p_{\mathrm{miss}}$ must be less than $0.3 \cdot \sqrts$,
the scalar sum of the transverse momenta of the particles in
the event must exceed 45 GeV, and
at least one pair of oppositely-charged tau candidates must be identified.

To select signal candidates, we use the two-tau likelihood of~\cite{pr183},
$\mathcal{L}_{\tau\tau}={\mathrm{\frac{P_1P_2}{P_1P_2+(1-P_1)(1-P_2)}}}$,
where ${\mathrm P}_i$ is the probability that the $i^{\mathrm th}$ tau candidate
originates from a real tau lepton.  This probability is calculated from
the shapes of the ANN output for signal and fake taus.  The distribution of
the ANN output for signal events was computed from Monte Carlo simulations.
This analysis has been improved by using for the fake taus
the distribution of the ANN output in
hadronic \Zo\ decay data collected in the calibration run at 
$\sqrt{s}\approx 91$~GeV, which
has a low fraction of events with real taus.  This estimation 
of the fake tau ANN distribution reduces the systematic
uncertainty on the fake tau rate.  To pass the selection, an event
must have $\mathcal{L}_{\tau\tau}$ of at least 0.10.

Next, the particles in the events
are subdivided into two tau candidates and two jets. 
A 2C kinematic fit is applied using total energy and momentum 
conservation constraints, where the tau momentum directions are taken 
from their visible decay products while leaving their energies free.
 The $\chi^2$-probability of the fit is required to be larger
than $10^{-5}$. 
In events where both taus are classified as one-prong decays,
the sum of the momenta
of the charged particles assigned to the tau decays
must be less than 80~GeV, in order to reduce backgrounds 
from $\ZZ\ra\mu^+\mu^-{\mathrm q{\overline q}}$ and 
$\ZZ\ra{\mathrm e^+e^-}{\mathrm q{\overline q}}$.

\begin{figure}[p]
\centerline{\epsfig{file=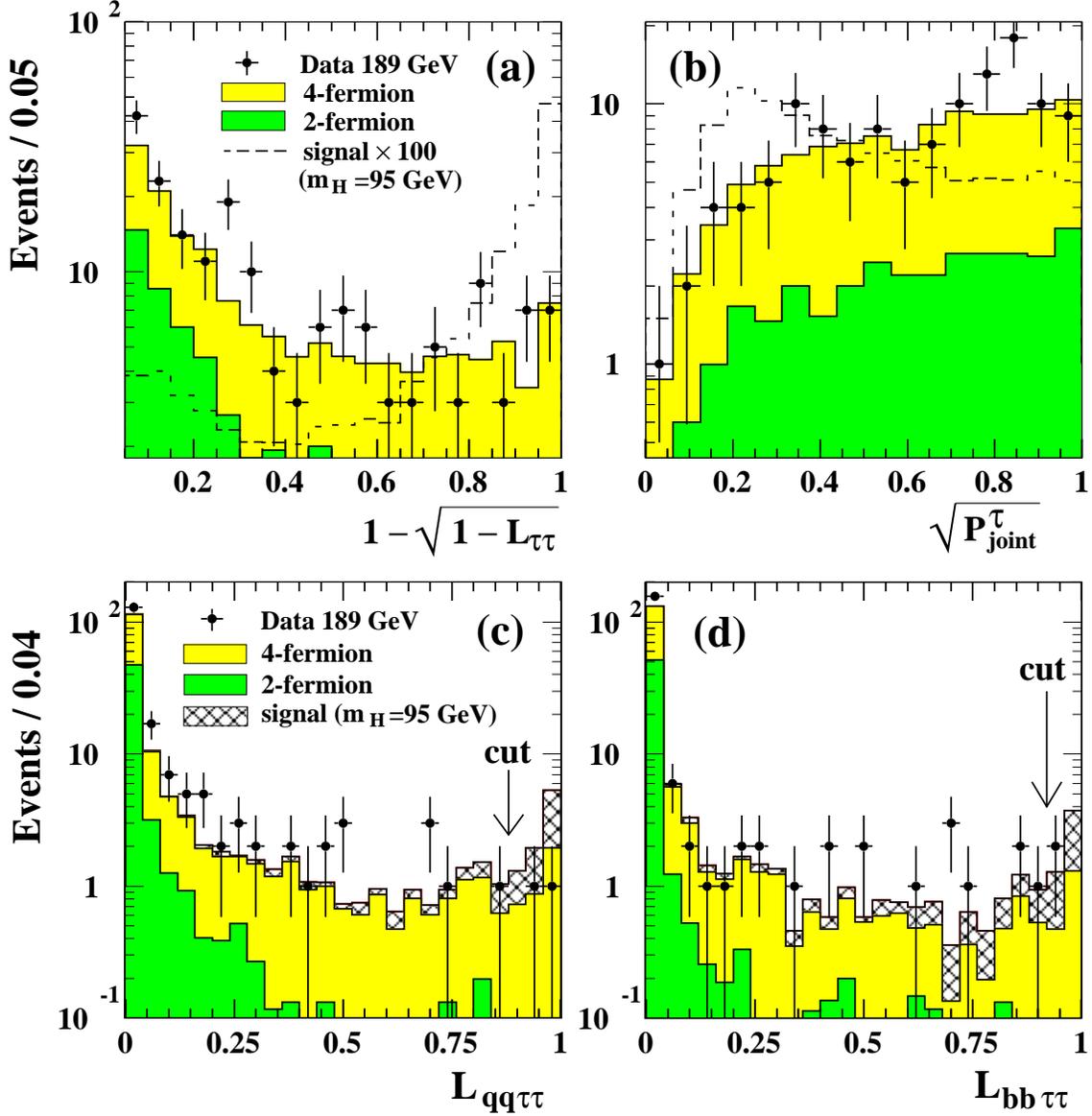,width=0.95\textwidth}}
\caption[]{\label{fig:tau}\sl
Tau channels: 
(a) distribution of the two-tau likelihood as input to the likelihood selection;
(b) distribution of the joint impact parameter significance 
as input to the likelihood selection;
(c) distribution of the ${\mathrm qq}\tau\tau$ likelihood;
(d) distribution of the ${\mathrm bb}\tau\tau$ likelihood; 
 OPAL data are
indicated by points with error bars, four-fermion backgrounds
by the light grey histograms, and two-fermion backgrounds by the
dark grey histograms.  Also shown is the contribution expected from
a 95~GeV Higgs boson (dashed histograms in (a) and (b), hatched histograms 
in (c) and (d)). In (a) and (b) the signal is scaled by a factor of 100.}
\end{figure}

The two final likelihoods, described in~\cite{pr183}, are then formed.
One, $\mathcal{L}( \bb\tautau )$,
is optimised for the $\ho\Zo\ra\bb\tautau$ final state 
and makes use of the b-tag of Section~\ref{sect:btag}.
The other, $\mathcal{L}( \qq\tautau )$, is optimised for the 
$\ho\Zo\ra\tautau\qq$ process and does not use b-tagging.
The following variables described in~\cite{pr183} 
are inputs to both likelihoods: 
\rvis = \evisn,
$\left| \cos \theta_{\mrm{miss}} \right|$,
$\mathcal{L}_{\tau\tau}$,
the logarithm of $y_{34}$,
the energy of the most energetic electron or muon identified in the event (if any),
the angles between each tau candidate and the nearest jet, and
the logarithm of the larger of the
two 3C kinematic fit probabilities, in which the additional constraint
comes from fixing either the
tau pair invariant mass or the jet pair invariant mass to the \Zo\ mass.
We have introduced a new variable which takes advantage of the finite lifetime
of the tau lepton, the impact parameter information of charged tracks
belonging to the tau candidate, combined in a 
joint-probability calculation~\cite{aleph_btag}.
In addition to the above variables,
the $\mathcal{L}( \bb\tautau )$ likelihood
uses the output of the b-tagging algorithm described in 
Section~\ref{sect:btag}.
An event is retained if
$\mathcal{L}( \bb\tautau )$ exceeds 0.92 
or $\mathcal{L}( \qq\tautau )$ exceeds 0.88.
Figure~\ref{fig:tau} shows the distributions of $\mathcal{L}_{\tau\tau}$, the
joint impact-parameter probability, and the two likelihoods,
$\mathcal{L}( \qq\tautau )$ and $\mathcal{L}( \bb\tautau )$, for the data,
simulated Standard Model backgrounds, and also for a simulated 95~GeV Higgs
signal.

The numbers of observed and expected events after each stage of the selection
are given in Table~\ref{tab:smflow}, together with 
the detection efficiency for a 95~GeV SM Higgs boson, which is estimated to be
(34.3 $\pm$ 1.1(stat.)$\pm$ 2.4(syst.))\% after the final selection requirement.
Three events survive the likelihood cut, to be compared to the expected
background of $4.0 \pm 0.5 (\mrm{stat}.) \pm 0.9 (\mrm{syst}.)$.
The systematic errors are evaluated as in~\cite{pr183}.

\label{sect:smqqttB}

 These results are confirmed by a separate analysis
 which uses a different technique to tag the tau candidates. 
 Events are reconstructed as four jets using the Durham
 algorithm
 and tau candidates are sought in the four jets using a likelihood 
 technique. The b-tagging algorithm of
 Section~\ref{sect:btag} is applied to the two remaining jets.
 This selection has a performance similar to that of the ANN method 
 for tau tagging described above. The efficiency for a 95~GeV Higgs signal
 is evaluated to be ($30.6 \pm 1.1({\mathrm stat.}) \pm 1.6({\mathrm syst.})$)\%.
 Six events are retained, two of which are selected also by the analysis above, 
 compared with the expected background of
 $3.0\pm 0.5$(stat.)$\pm 0.6$(syst.) events. 

\subsection{The Electron and Muon Channels}
\label{sect:smlept}
\begin{figure}[p]
\centerline{\epsfig{file=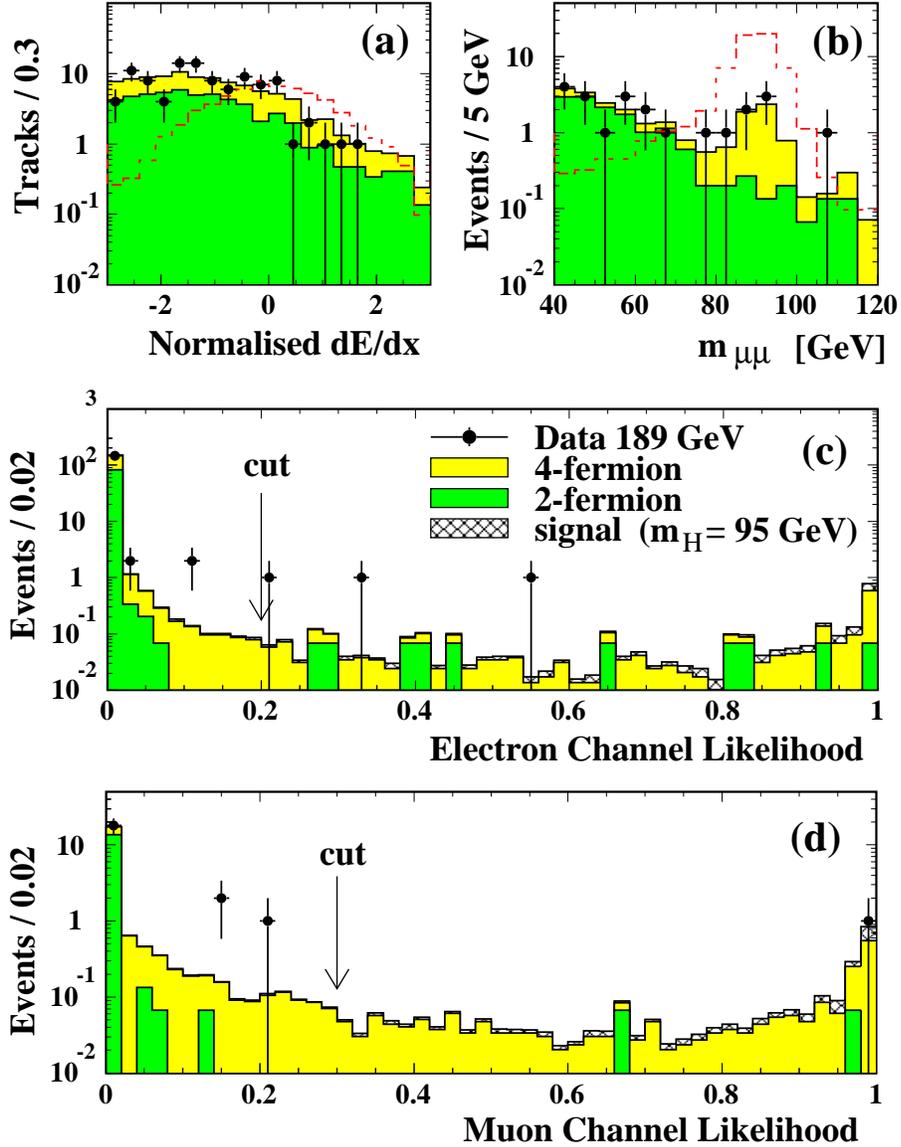,width=0.75\textwidth}}
\caption[]{\label{fig:lepton} Electron and Muon channels:
(a) The distribution of the normalised d$E$/d$x$, the difference between the
ionisation energy loss of a track and that expected for an electron, divided by
the measurement error, for tracks in the high-energy data.
(b) Reconstructed invariant mass of identified muon pairs, where the signal is expected
to contribute near $m_{\mu\mu}=m_{\mathrm Z^0}$.  (c) The electron channel likelihood 
distribution, and (d) the muon channel likelihood distribution.
OPAL data are
indicated by points with error bars, four-fermion backgrounds
by the light grey histograms, and two-fermion backgrounds by the
darker grey histograms. The contribution from a 95~GeV Higgs boson signal
is shown with dashed histograms in (a) and (b) and with hatched histograms 
in (c) and (d). It is scaled by a factor of 100 in (a) and (b).
}
\end{figure}

The preselection for this analysis is
intended to enhance the \leplep\qq\ topology, where $\ell=$e or $\mu$.
The requirements are listed below.
\begin{enumerate}
\item The event must have at least six charged tracks, $y_{34} > 10^{-4}$,
$|P^z_{\mathrm{vis}}| < (E_{\mathrm vis}-0.5\sqrt{s})$ and
$E_{\mathrm vis} > 0.6\sqrt{s}$.
\item At least one pair of oppositely charged leptons of the
same flavour (e or $\mu$) must be identified as described in~\cite{smpaper172}.
Figure~\ref{fig:lepton}(a) shows the distribution of the 
normalized d$E$/d$x$, an important ingredient
of the electron identification.
\item The events are reconstructed as two leptons and two jets.  In the case of
the muon
channel, a 4C kinematic fit, requiring total energy and momentum conservation,
is applied to improve the mass resolution of the muon pair and is
required to yield a $\chi^2$ probability larger than 10$^{-5}$.
The invariant mass of the lepton pair is required to be larger than 40 GeV\@.
The distribution of reconstructed dimuon masses is shown in Figure~\ref{fig:lepton}(b).
\end{enumerate}

Two likelihoods, one based on kinematic variables, $\cal K$, and one for b-tagging
for the two jets of hadrons, ${\cal B}_{\mathrm 2jet}$, are calculated as
described in~\cite{pr183}.
The final signal likelihood is computed using $\cal K$ and 
${\cal B}_{\mathrm 2jet}$ as inputs.  The combined likelihood is required to
exceed 0.2 for the electron channel and 0.3 for the muon channel.
Distributions of these likelihoods are shown in Figure~\ref{fig:lepton}.
The signal selection efficiency for a
95~GeV SM Higgs boson
is (55.0$\pm$0.9(stat.)$\pm$1.1(syst.))\%  for the electron channel, and
(64.9$\pm$0.9(stat.)$\pm$0.9(syst.))\% for the muon channel. 
The numbers of observed and expected events after each stage of the selection
are given in Table~\ref{tab:smflow}, together with 
the detection efficiency for a 95~GeV SM Higgs boson. 
The selection retains three events in the electron channel and one in the
muon channel. The total background
expectation is 2.6$\pm$0.2(stat.)$\pm$0.5(syst.) events in the electron channel
and 2.1$\pm$0.1(stat.)$\pm$0.4(syst.) events in the muon channel.
%

%

\begin{table}[htbp]
\vspace*{-1.0cm}
\begin{center}
\begin{tabular}{|c||r||r||r|r||c|}\hline
Cut & Data & Total & q\=q($\gamma$) & 4-fermi. & Efficiency (\%)\\
    &      &   bkg.&    bkg.        &   bkg.   & $\mh=95$~GeV \\\hline\hline
\multicolumn{6}{|c|}{Four-jet Channel ~~ 172.1 pb$^{-1}$}     \\ \hline
    (1)  &  18701    &  18120  &    14716 &  3200  &   99.9 \\
    (2)  &   6242    &   6183  &     4215 &  1954  &   95.6 \\
    (3)  &   1955    &   1891  &      538 &  1353  &   91.2 \\
    (4)  &   1927    &   1864  &      513 &  1351  &   90.1 \\
    (5)  &   1729    &   1668  &      436 &  1231  &   89.8 \\
    (6)  &   1555    &   1506  &      378 &  1128  &   88.4 \\\hline
${\cal L}^{\mathrm HZ}$& 24  &   19.9  &      4.9 &  15.0  &  47.0 \\\hline\hline
\multicolumn{6}{|c|}{Missing-energy Channel ~~ 171.4 pb$^{-1}$}  \\ \hline
(1) &  4267 & 4252 & 3201  & 1029  & 77.9\\
(2) &  1032 & 1062 &  341  &  717  & 74.1\\
(3) &   981 & 1016 &  328  &  684  & 73.1\\
(4) &   650 &  684 &   56  &  628  & 62.4\\
(5) &   184 &  175 &   21  &  154  & 59.2\\
(6) &   111 &  101 &   18  &   83  & 57.4\\\hline
${\cal L}^{\mathrm HZ}$&10& 6.9 & 1.1& 5.7 &  35.4 \\\hline\hline
\multicolumn{6}{|c|}{Tau Channel ~~ 168.7 pb$^{-1}$}  \\ \hline
Pre-sel             &  4652   & 4584  &  2809 & 1767  &    80.4\\
$\mathcal{L}_{\tau\tau}$&733   &  693  &   100 &  590  &    62.3\\
2C fit                 &201   &  160  &    56 &  104  &    50.3\\
1-prong E sum          &185   &  156  &    55 &  101  &    50.0\\\hline
Final $\mathcal{L}$    &  3   &  4.0  &   0.1 &  3.9  &    34.3\\\hline\hline
\multicolumn{6}{|c|}{Electron Channel 172.1 pb$^{-1}$} \\\hline\hline
    (1)        & 8371 & 8587 & 6219 & 2367 & 91.0 \\
    (2)        &  521 &  508 &  405 &  102 & 86.4 \\
    (3)        &  152 &  153 &   84 &   69 & 75.5 \\\hline
$\cal{L}^{\mathrm HZ}$ & 3&  2.6 & 0.7 & 1.9 & 55.0\\\hline\hline
\multicolumn{6}{|c|}{Muon Channel 169.4 pb$^{-1}$} \\\hline\hline
      (1)        & 8232 & 8452 & 6122 & 2330 &91.0\\
      (2)        &  103 &  100 &   71 &   29 &78.2\\
      (3)        &   22 &   22 &   14 &    8 &75.9\\\hline
$\cal{L}^{\mathrm HZ}$ &  1 & 2.1 & 0.1 &2.0 &64.9\\\hline
\end{tabular}
\caption{\label{tab:smflow}\sl
  The \ho\Zo\  channels:
  the numbers of events after each cut for 
  the data and the expected background, normalised to the data luminosity.
  The two-photon background, not shown separately, is included in the total
  background.
  The last column shows the detection efficiencies, for
  \ho\ra~\bb\ in the four-jet channel, for \ho\ra~all in
  the missing-energy, electron, and muon channels, and for
  \Zo\ho\ra\tautau(\ho\ra~all) or \Zo\ho\ra\qq\tautau\ in the tau channel, 
  for a Higgs boson mass of 95\,GeV with Standard Model branching fractions.}
\end{center}
\end{table}

\subsection{Search for the MSSM Process ${\protect \boldmath \ho\Zo}$ 
with ${\protect \boldmath \ho\ra\Ao\Ao}$ \label{sect:hzaa}} 

If $2\mA\leq\mh$, the decay 
\ho\ra\Ao\Ao\ 
is kinematically allowed and is the dominant decay of the
\ho\ in parts of MSSM parameter space.
Dedicated, optimised searches for \ho\Zo\ra\Ao\Ao\Zo\
have not been performed. Instead, we
evaluate the sensitivity of the \ho\Zo\ searches
for processes with \ho\ra\Ao\Ao\ by studying the efficiencies of the
selections using a {\ho\Zo\ra\Ao\Ao\Zo\ra\bb\bb\Zo}
Monte Carlo.   Non-\bb\ decays of the \Ao\ are not
considered.  The efficiencies of the \ho\Zo\ selections
for the process \ho\Zo\ra\Ao\Ao\Zo\ 
with $\mA = 20$~GeV and $\mh = 70$~GeV are summarised
in Table~\ref{tab:hAA}. 
In general, the efficiency is found to increase with the \ho\ mass.
This procedure has the advantage of simplifying the statistical
treatment and is sufficiently powerful to exclude regions in which the decay
\ho\ra\Ao\Ao\ra\bb\bb\ dominates. 
\begin{table}[htbp]
\begin{center}
\begin{tabular}{|l|l|c|}
\hline
SM search & applied to the process & Efficiency (\%) \\  
\hline\hline
four jet        &(\Ao\Ao\ra\bb\bb)(\Zo\ra\qq)     & 20 \\
missing energy  &(\Ao\Ao\ra\bb\bb)(\Zo\ra\nn)     & 32 \\
tau lepton      &(\Ao\Ao\ra\bb\bb)(\Zo\ra\tautau) & 38 \\
electron        &(\Ao\Ao\ra\bb\bb)(\Zo\ra\ee)     & 60 \\
muon            &(\Ao\Ao\ra\bb\bb)(\Zo\ra\mm)     & 73 \\
\hline
\end{tabular}
\caption[]{\label{tab:hAA}\sl
     Signal detection efficiencies for the searches for the SM Higgs boson, 
     applied to the processes with \ho\ra\Ao\Ao\ followed by \Ao\ra\bb. 
     The efficiencies are quoted for $\mh=70$~GeV and
     $\mA=20$~GeV.  The statistical errors due to the limited sizes of the
     Monte Carlo samples are 1--4\%.  The backgrounds and candidates for each
     channel are listed in Table~\ref{tab:smflow}.
}
\end{center}
\end{table}
The expected backgrounds and the selected candidates are
the same as for the results described above,
since the same selections are applied.

\section{The ${\protect \boldmath \Ao\ho}$ search channels}
\label{sect:hasearches}
We search for the MSSM process of associated production $\ee\ra\Ao\ho$
followed by the decays
\Ao\ho\ra\bb\bb\ and \Ao\ho\ra\bb\tautau. 
If \ho\ra\Ao\Ao\ is kinematically allowed, the process
\ee\ra\ho\Ao\ra\Ao\Ao\Ao\ is also searched for in the \bb\bb\bb\ final
state.  These selections differ from those for \Zo\ho\ because
the \Zo\ mass constraint is no longer applicable.  The selections
therefore have been optimised to reject backgrounds differing
kinematically from those important to the \Zo\ho\ searches.

\subsection{The ${\protect \boldmath \Ao\ho\ra\bb\bb}$ Final State}
\label{sect:habbbb}
\begin{figure}[p]
\centerline{\epsfig{file=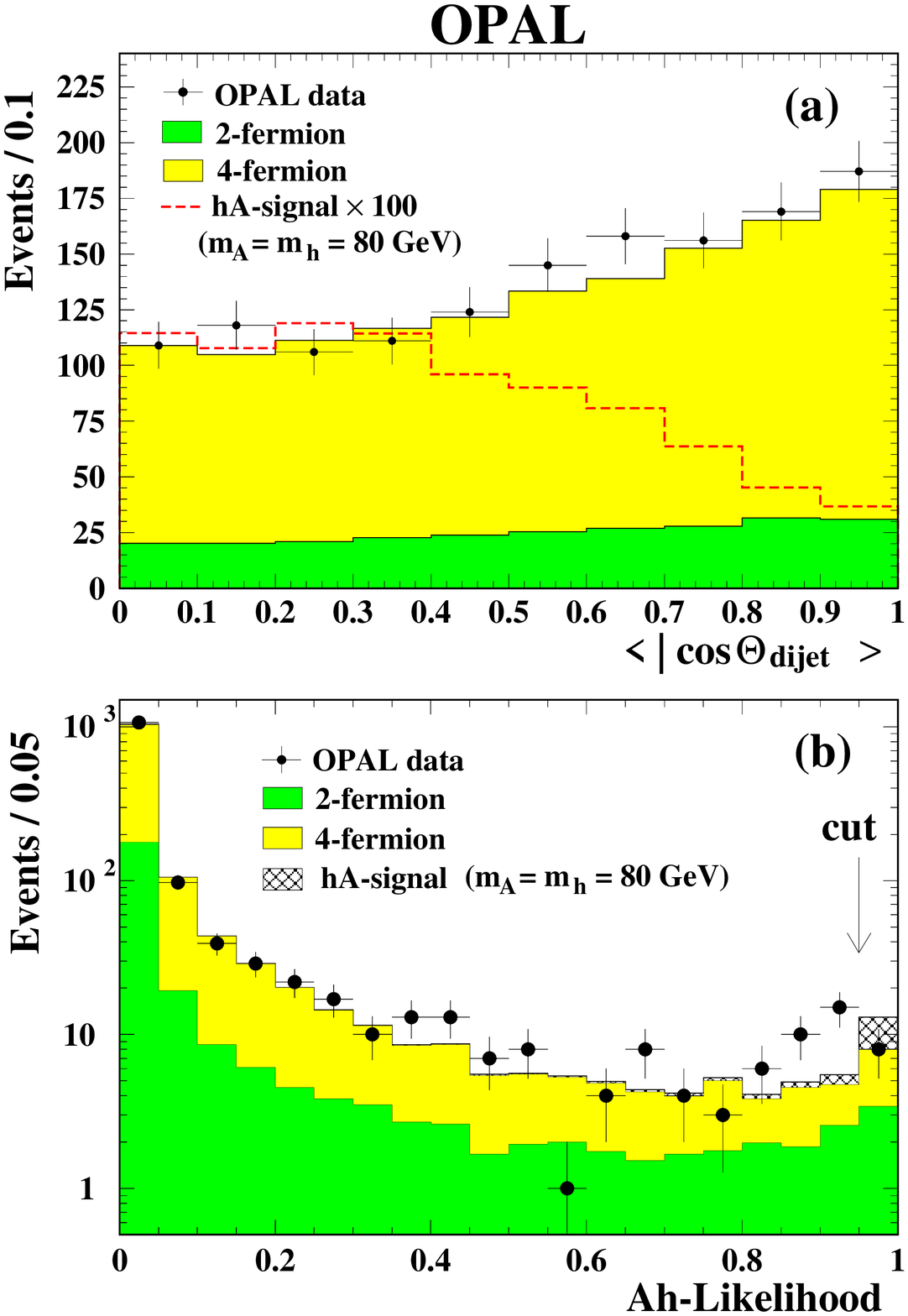, width=0.8\textwidth}}
\caption[]{\label{fig:bbbb}\sl
(a) The distribution of $\langle|\cos\Theta_{\mathrm dijet}|\rangle$
for the \Ao\ho\ra\bb\bb\ channel.
(b) The distribution of the selection likelihood in the \Ao\ho\ra\bb\bb\ channel.
OPAL data are
indicated by points with error bars, four-fermion backgrounds
by the light grey histograms, and two-fermion backgrounds by the
dark grey histograms.  Also shown is the contribution expected from
a Higgs boson signal with $m_{\mathrm h}=m_{\mathrm A}=80$~GeV
(other MSSM parameters are set to their ``benchmark'' values with
maximal scalar top mixing, as described in Section~\ref{sect:mssmlimit})
as dashed (a) and hatched (b) histograms. 
In (a) the signal is scaled by a factor of 100.
}
\end{figure}

The same preselection requirements are applied as described in~\cite{pr183}. 
These are: 
\begin{enumerate}
\item[1--3.]  The requirements 1--3 of the \ho\Zo\ four-jet analysis
              of Section~\ref{sect:sm4jet} are used.
\item[4.] The $C$-parameter must be larger than 0.45.
\item[5.] Each of the four jets must contain at least six energy-flow objects
and at least one charged track.
\item[6.] The $\chi^{2}$ probability of a 4C fit, which requires energy and momentum
conservation, must be larger than $10^{-5}$.
\end{enumerate}
For events passing the preselection, a likelihood technique
is applied as described in~\cite{pr183}.
Seven input variables are used. Six of these variables are the
same as those used in~\cite{pr183}: the four b-tagging discriminants \Dbi, one
for each jet (see Section~\ref{sect:btag}),
$y_{34}$ in the Durham scheme and the event thrust. The seventh variable is
$\langle|\cos\Theta_{\mathrm{dijet}}|\rangle$, the average
absolute value of the cosines of the
polar angles of the momenta of the two di-jet systems for the combination yielding the 
smallest difference in the two di-jet invariant masses after the 
4C kinematic fit.
This last variable replaces the mean $|\cos{\theta}_{\mathrm{jet}}|$ of the four jets used 
in~\cite{pr183} because it provides better discrimination
against the W$^{+}$W$^{-}$ background.
The distribution of $\langle|\cos\Theta_{\mathrm{dijet}}|\rangle$ is shown in 
Figure~\ref{fig:bbbb}, along with the final likelihood discriminant
${\cal L}^{\mathrm{Ah}}$.

Candidate events are selected by requiring 
${\cal L}^{\mathrm{Ah}} > \mbox{0.95}$. 
Table~\ref{tab:ahflow} shows the number of selected events together
with the expectation from background processes and the signal
selection efficiency for \mA = \mh = 80~GeV,
after each cut in the
preselection and after the final cut on ${\cal L}^{\mathrm{Ah}}$. 
The detection efficiency for a Higgs signal with masses
\mA = \mh = 80~GeV is estimated to be (48.4$\pm$0.7(stat.)$\pm$3.9(syst.))\%.
Eight candidate events are observed in the data, consistent with
8.0$\pm$0.5(stat.)$\pm$1.4(syst.) events expected 
from SM background processes.
Six of the candidate events are common to those found in the
SM four-jet channel of Section~\ref{sect:sm4jet}.
The systematic uncertainties on the
signal selection efficiencies and background estimates 
were determined using the same methods as described
in~\cite{pr183}.

\begin{figure}[p]
\centerline{\epsfig{file=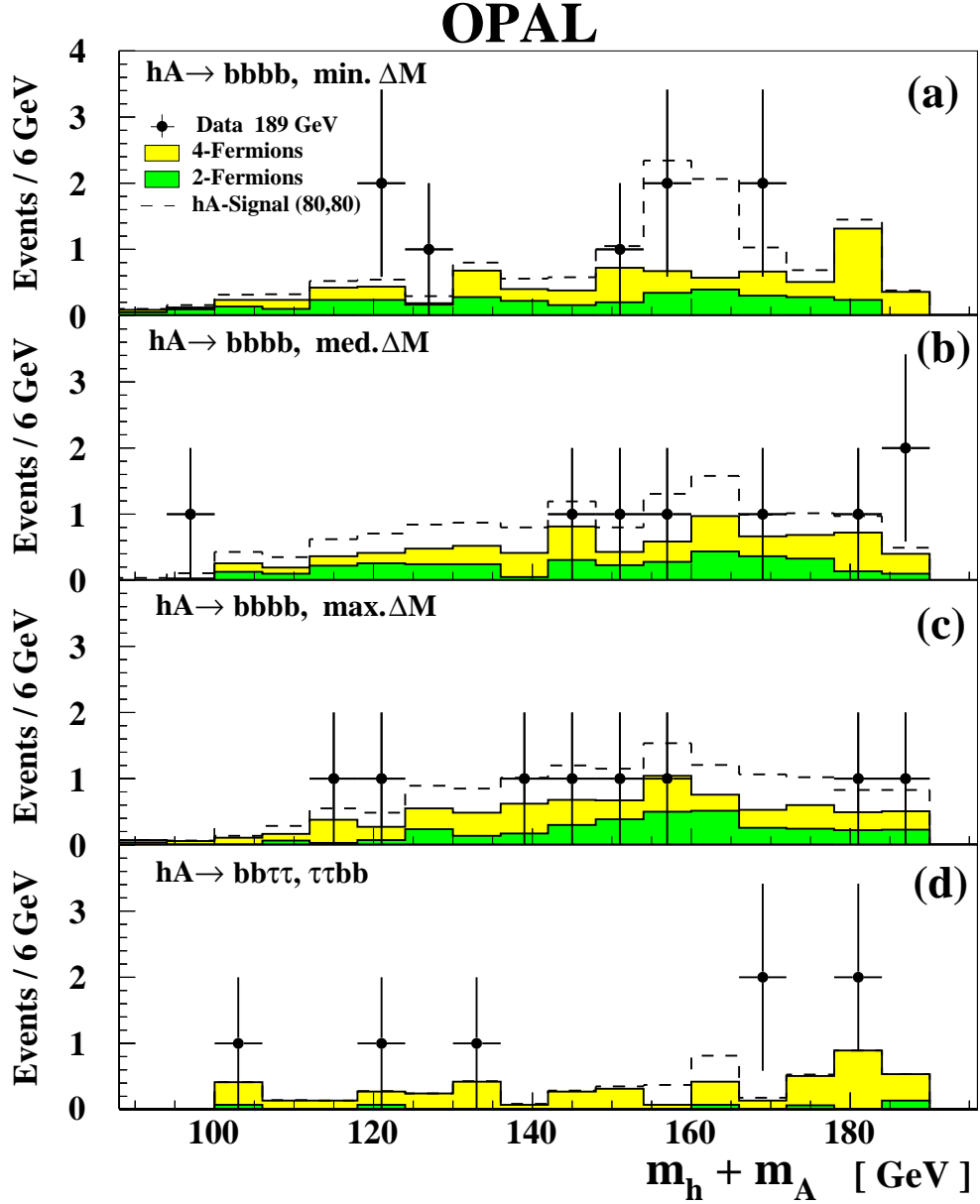,width=1.0\textwidth}}
\caption[]{\label{fig:ahmass}\sl The sum of the reconstructed Higgs masses, 
$m_{\mathrm A}^{\mathrm rec} + m_{\mathrm h}^{\mathrm rec}$, for
        (a) the Ah$\rightarrow$4b channel with the di-jet pairing combination 
            which yields the smallest
            mass difference,
            $\Delta{\mathrm{M}} \equiv |m_{\mathrm A}^{\mathrm rec} - m_{\mathrm h}^{\mathrm rec}|$,
        (b) the Ah$\rightarrow$4b channel with the medium $\Delta$M combination,
        (c) the Ah$\rightarrow$4b channel with the maximum $\Delta$M combination, and
        (d) the Ah$\rightarrow \tautau \bb , \bb \tautau $ channel.
OPAL data are
indicated by points with error bars, four-fermion backgrounds
by the light grey histograms, and two-fermion backgrounds by the
darker grey histograms. 
Also shown as dashed histograms are the contributions expected from
a Higgs boson signal with $m_{\mathrm h}=m_{\mathrm A}=80$~GeV
(other MSSM parameters are set to their ``benchmark'' values with
maximal scalar top mixing, as described in Section~\ref{sect:mssmlimit}).
}
\end{figure}

Candidate Higgs masses are calculated
from the measured jet momenta using the 4C fit.
Figure~\ref{fig:ahmass}(a)-(c) shows the distribution of the sum of the
reconstructed Higgs masses, $M_{\mathrm{sum}} \equiv m_{\mathrm h}^{\mathrm rec}+m_{\mathrm A}^{\mathrm rec}$, for all 
three possible di-jet pairings,
separately according to the reconstructed Higgs mass difference, 
$\Delta{\mathrm{M}} \equiv |m_{\mathrm A}^{\mathrm rec} - m_{\mathrm h}^{\mathrm rec}|$. 
The resolution on the mass sum, $M_{\mathrm{sum}}$, is 
estimated to be approximately 3~GeV for $M=150$ GeV.  For $m_{\mathrm h}=m_{\mathrm A}$,
68\% of the events have a reconstructed mass difference 
$\Delta{\mathrm{M}}$ of less than 13~GeV.
Since the four jets can be combined in three ways, and since
the \ho\ and \Ao\ cannot be distinguished, each candidate event enters at
six points in the (\mh ,\mA) plane.

\subsection{The ${\protect \boldmath \Ao\ho\ra\bb\tautau}$ Final State}

The \Ao\ho\ra\bb\tautau\ final state, where either the \Ao\ or the
\ho\ decays into a tau pair, has been searched for using the
same technique as used in the SM tau channels described above.
The final likelihood selection has been optimised for the
MSSM process.

The following variables are used as inputs to the calculation of
the likelihood, $\mathcal{L}_{\mrm{hA}}$: 
\rvis,
$\left| \cos \theta_{\mrm{miss}} \right|$, $\mathcal{L}_{\tau\tau}$,
the logarithm of $y_{34}$,
the energy of the most energetic identified electron or muon,
the tau-track lifetime information as described
in Section~\ref{sect:smbbtt},
the average of the absolute values of the cosines of the polar angles of the
reconstructed \ho\  and \Ao, which gives an estimate of the production angle of the
two Higgs bosons, and the
outputs of the b-tagging algorithm described in Section~\ref{sect:btag} for
the two hadronic jets in the event.

Candidate events are selected by requiring the final likelihood
value to exceed 0.64.  Table~\ref{tab:ahflow} shows the number of
selected events, the efficiency for $\mh =\mA = 80$~GeV and the
background estimation, after each stage of the selection.
The efficiency for a signal with $\mh =\mA = 80$~GeV
is estimated to be (45.3$\pm$1.5(stat.)$\pm$2.3(syst.))\%.
Seven candidates are observed in the data,
two of which are shared with the SM analysis of
Section~\ref{sect:smbbtt}.
The number of events expected from SM background processes is
$4.9 \pm 0.6$(stat.)$\pm 1.6$(syst.).
The distribution of the sum of the reconstructed Higgs masses,
$M_{\mathrm h}^{\mathrm rec}+M_{\mathrm A}^{\mathrm rec}$, 
is shown in Figure~\ref{fig:ahmass}(d). 
Since the \Ao\ and \ho\ cannot be distinguished,
each selected event has two interpretations in the (\mh ,~\mA) plane.
Systematic uncertainties on backgrounds and
efficiencies are evaluated as in Section~\ref{sect:smbbtt}.

The alternative jet-based analysis, described in 
Section~\ref{sect:smbbtt}, is also applied in this search channel. This
procedure gives an overall efficiency 
of (39.1$\pm$1.5)\% for $\mh =\mA = 80$~GeV, similar to that obtained
above with the ANN tau identification, and the efficiency remains similar
over the range of kinematically allowed values of \mh\ and \mA.
The choice of the ANN analysis has been made to optimise the expected limits.
Five events, two of which are
shared with the ANN analysis, meet the selection requirements of the
alternative analysis, 
consistent with the expectation of  
$4.8 \pm 0.6$(stat.)$\pm 0.9$(syst.) from SM
background processes.

\subsection{The ${\protect \boldmath 
\Ao\ho\ra\Ao\Ao\Ao\ra\bb\bb\bb}$ Final State}\label{section:ah6b}
Signal events in this decay mode are characterised by a large number
of jets containing b-flavoured hadrons.
Backgrounds are reduced through cuts on kinematic variables and
by b-tagging.
Signal events satisfy $2\mA\leq\mh$, and 
this analysis uses an ANN to optimise the sensitivity
over the allowed mass region in the (\mh,\mA) plane. 

The analysis begins with a set of preselection requirements.
Events must be selected as hadronic final states~\cite{l2mh}. 
The polar angle of the thrust axis, $\theta_{T}$, must
satisfy $|\cos\theta_{T}|\le 0.9$.
The charged tracks and clusters are grouped into six jets
using the Durham~\cite{durham} algorithm.
Each jet is required to have at least one charged track and one 
electromagnetic cluster.  Events must have at least 20
good quality charged tracks and at least 20 good quality electromagnetic
calorimeter clusters, where the quality requirements are those used
in~\cite{higgsold}.  To suppress the background from \Zgs, the value of
$y_{34}$ is required
to be larger than 0.0005 and the $C$-parameter to be larger than 0.0075.

After the preselection,
candidates are selected using two neural networks applied sequentially, one
combining kinematic and topological variables, 
and one for the b-tagging variables.
The kinematic characteristics of signal events depend strongly on $m_{\mathrm h}$
and $m_{\mathrm A}$, and in order to maintain good sensitivity over a broad
range of these masses, 14 variables with complementary discriminating roles
are used as inputs for the first network. These are the
event thrust, sphericity, oblateness, $C$ and $D$ parameters~\cite{cpar},
the $2^{\mathrm{nd}}$ through the $6^{\mathrm{th}}$ 
normalised Fox-Wolfram~\cite{FOX} moments, and
$y_{23}$, $y_{34}$, $y_{45}$ and $y_{56}$ for the Durham scheme.
Events were required to have a kinematic network output greater than 0.68.

\begin{figure}[p]
\centerline{\epsfig{file=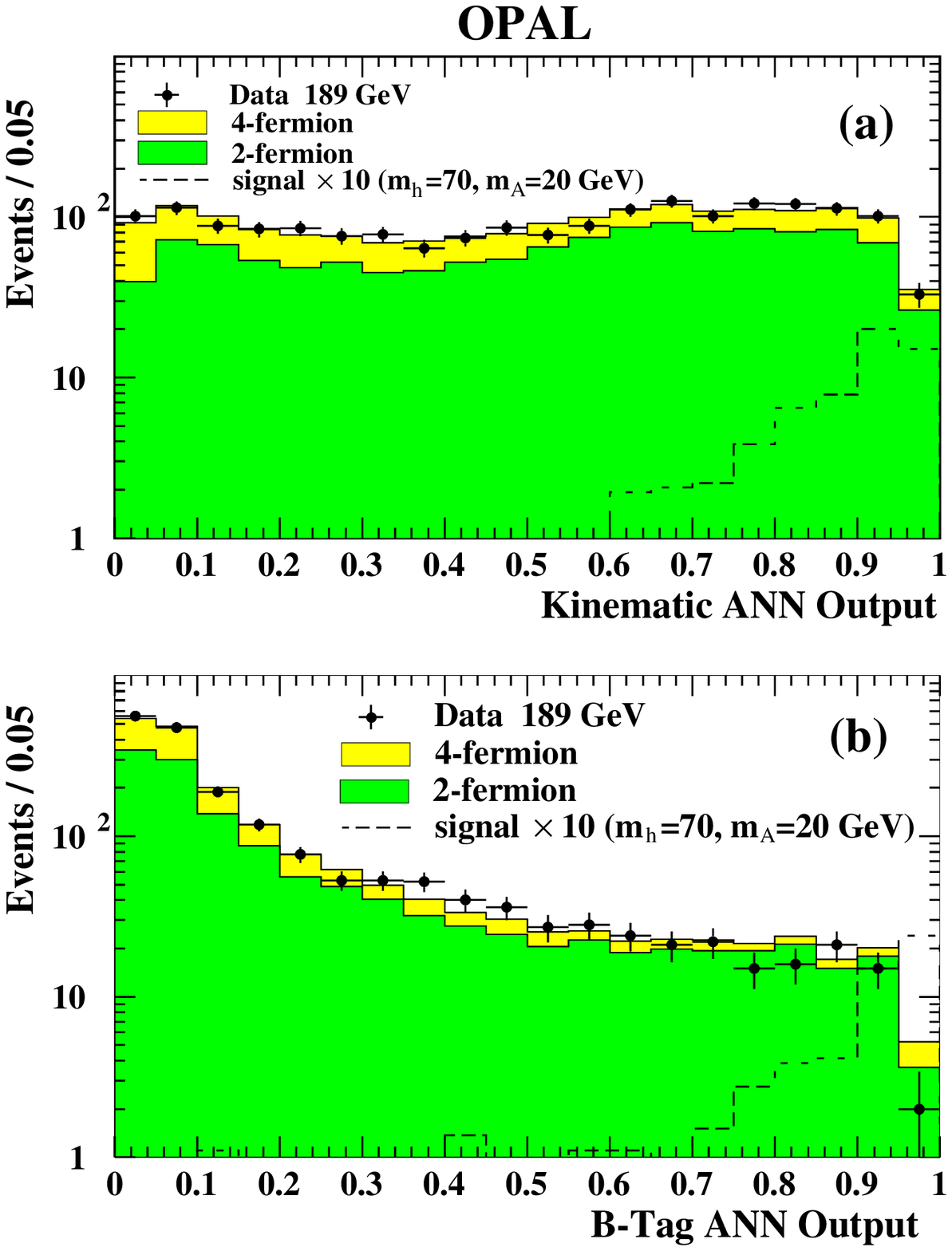,height=0.80\textheight}}
\caption[]{\label{fig:6b}\sl
$\ee\ra\Ao\ho\ra\Ao\Ao\Ao\ra\bb\bb\bb$ channel: 
The distribution of the ANNs with kinematic variables (a) and with 
b-tagging variables (b), as described in the text. The light grey histograms
indicate Standard Model $\qq$ background, and the dark grey histograms
show the distributions of Standard Model four-fermion background.
The dashed histograms show the distribution from a signal Monte Carlo 
with Higgs boson masses of
$m_{\mathrm h}$=70~GeV and $m_{\mathrm A}$=20~GeV (other MSSM parameters are
 set to their ``benchmark'' values with
maximal scalar top mixing, as described in Section~\ref{sect:mssmlimit})
scaled by a factor of 10.
}
\end{figure}

The final selection was made with an ANN optimised for b-tagging.
If the \Ao\ is light, the two jets originating from
the \Ao\ decay may be observed as a single jet; events then may seem to have
only three jets.  The events are therefore reconstructed both as three and six jets
and a 15-input neural network has been trained using the b-tag variables
described in Section~\ref{sect:btag}, 
from the three and six jets in the event. 
Events are required to have a b-tagging network output greater than 0.92.
The distributions of the two network outputs
are shown in Figure~\ref{fig:6b} for events passing the preselection.


Five events pass all selection requirements~(Table~\ref{tab:ahflow}), consistent with
an expected background of 8.7$\pm$1.0(stat.)$\pm$2.5(syst.) events.
The signal efficiency for $\mA$=20 and $\mh$=70 GeV
is (45.4$\pm$2.2(stat.)$\pm$4.3(syst.))\%.
The systematic errors are evaluated similarly
to those for the other channels~\cite{pr183}.
The dominant contribution to the systematic error on the signal and background
efficiencies comes from the uncertainty related to the b-tagging.
The main components arise from uncertainties in 
the tracking performance of the detector and the
b-hadron decay multiplicity~\cite{bmul}.

\begin{table}[ht]
\begin{center}
\begin{tabular}{|c||r||r||r|r||c|}\hline
Cut & Data & Total bkg. & q\=q($\gamma$) & 4-fermi. & Efficiency (\%)\\\hline\hline
\multicolumn{5}{|c||}{Ah-4b Channel ~~ 172.1 pb$^{-1}$}&$\mA = \mh =80$ GeV\\\hline
(1) &  18701 &  18120  & 14716 &   3200  &   99.9 \\
(2) &   6242 &   6183  &  4215 &   1954  &   95.6 \\
(3) &   1955 &   1891  &   538 &   1353  &   91.2 \\
(4)&    1668&    1606&     316&    1290&     83.3\\
(5)&    1464&    1402&     273&    1128&     83.3\\
(6)&    1382&    1335&     251&    1084&     81.3\\\hline
${\cal L}^{Ah}>0.95$ & 8& 8.0&  3.4& 4.6&  48.4\\\hline\hline
\multicolumn{5}{|c||}{Ah-tau Channel ~~ 168.7 pb$^{-1}$} & $\mA = \mh =80$ GeV \\ \hline
Pre-sel             &  4652   & 4584  &  2809 & 1767  &    83.8\\
$\mathcal{L}_{\tau\tau}$&733  &  693  &   100 &  590  &    69.4\\
2C fit                 &201   &  160  &    56 &  104  &    55.2\\
1-prong E sum          &185   &  156  &    55 &  101  &    54.8\\\hline
Final $\mathcal{L}$    &  7   &  4.9  &   0.4 &  4.5  &    45.3\\\hline\hline
\multicolumn{5}{|c||}{Ah-6b Channel ~~ 172.1 pb$^{-1}$} & ($\mh , \mA$)=(70,20) GeV \\ \hline
Pre-sel  & 1841  &  1740            & 1205   & 535 & 90.4 \\
Kin-ANN  &  649  &   591            &  438   & 153 & 82.6 \\\hline
btag-ANN &    5  &   8.7            &  7.7   & 1.0 & 45.4 \\
\hline
\end{tabular}
\end{center}
\caption{\label{tab:ahflow}\sl
         The \Ao\ho\  channels:
         effect of the cuts on data and the simulated
         background, normalised to the integrated luminosity of the data. 
         The two-photon backgrounds, not shown separately, are included in
         the total background. 
         The signal efficiencies are given in the last column for
         \mh=\mA=80~GeV in the \ho\Ao\ra\bb\bb\  and the tau channels and 
         for $\mh=70$~GeV and $\mA=20$~GeV in 
         the \ho\Ao\ra\Ao\Ao\Ao\ra\bb\bb\bb\  channel.
}
\end{table}

\section{Limits}

\subsection{Mass limit for the SM Higgs boson}
\begin{figure}[p]
\centerline{\epsfig{file=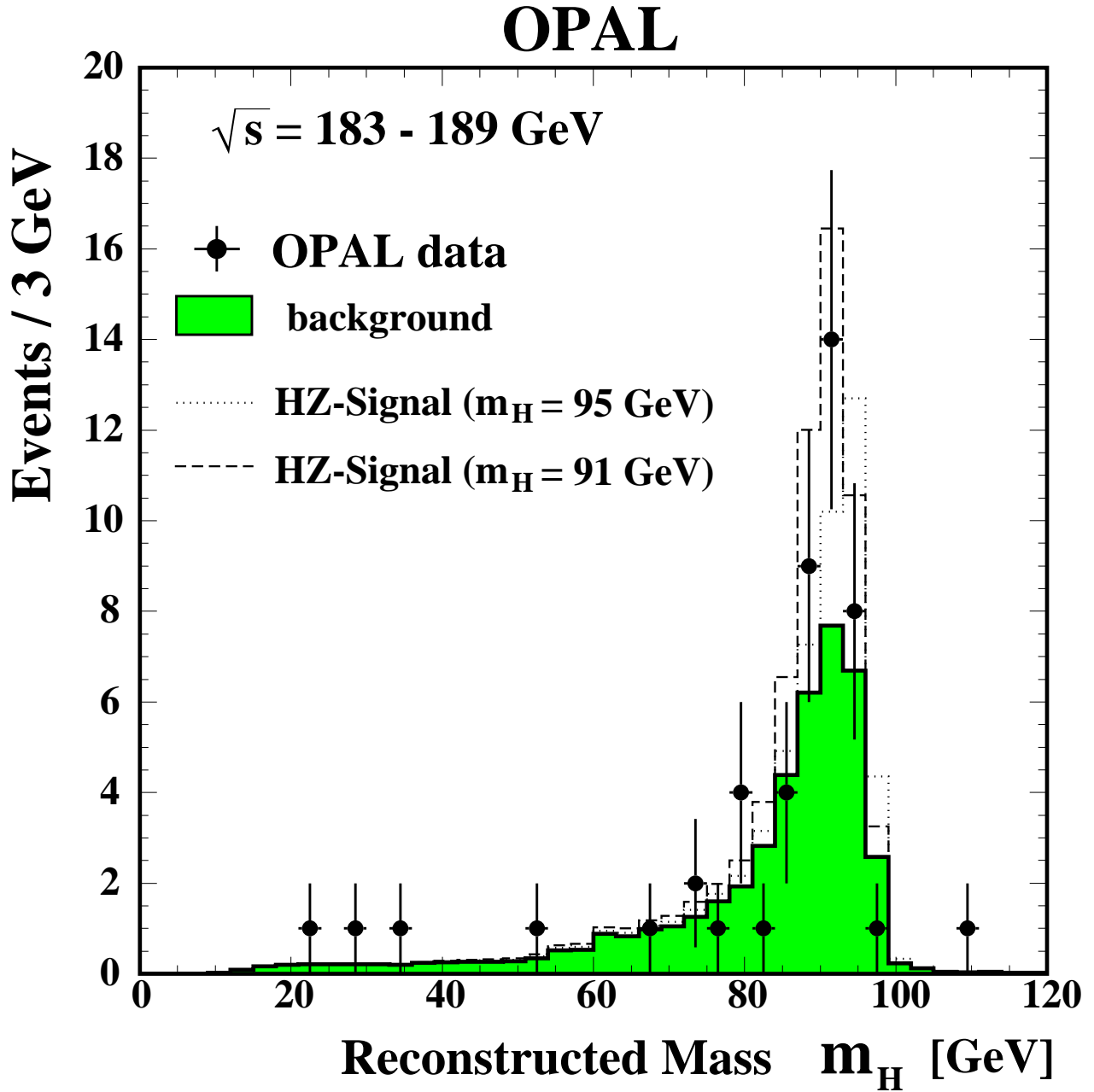,width=17cm}}
\caption[]{\label{fig:smmass}\sl
          The mass distribution for the selected events in all \ho\Zo\ 
          channels combined (points with error bars), 
          and the expected background.  The data and expected backgrounds
          have been combined for the $\sqrts=183$~GeV
          and $\sqrts=189$~GeV samples.  The expected mass distribution assuming the
          production of the SM Higgs boson with a mass of 91 GeV (95 GeV) is added on top of
          the background and shown with a dashed (dotted) histogram. 
}
\end{figure}

We summarise the efficiencies and numbers of expected 
SM Higgs signal and background events in Table~\ref{tab:smsummary},
separately for 189 GeV and with the 183 GeV sample~\cite{pr183} added.
A total of 41 events at 189~GeV pass the selections while 
$35.4\pm 1.1({\mathrm stat.}) \pm 3.2({\mathrm syst.})$ 
events are expected from the SM background processes.
The reconstructed masses of the candidate events
and the distribution expected for the combined SM backgrounds
for the searches presented here, added to those
from our $\sqrts \approx 183$~GeV searches~\cite{pr183},
are shown in Figure~\ref{fig:smmass}.  A slight excess of events peaks at the
\Zo\ mass, which is consistent with a statistical fluctuation of the dominant \Zo\Zo\
background.

\begin{figure}[p]
\vspace*{-0.7cm}
\centerline{\epsfig{file=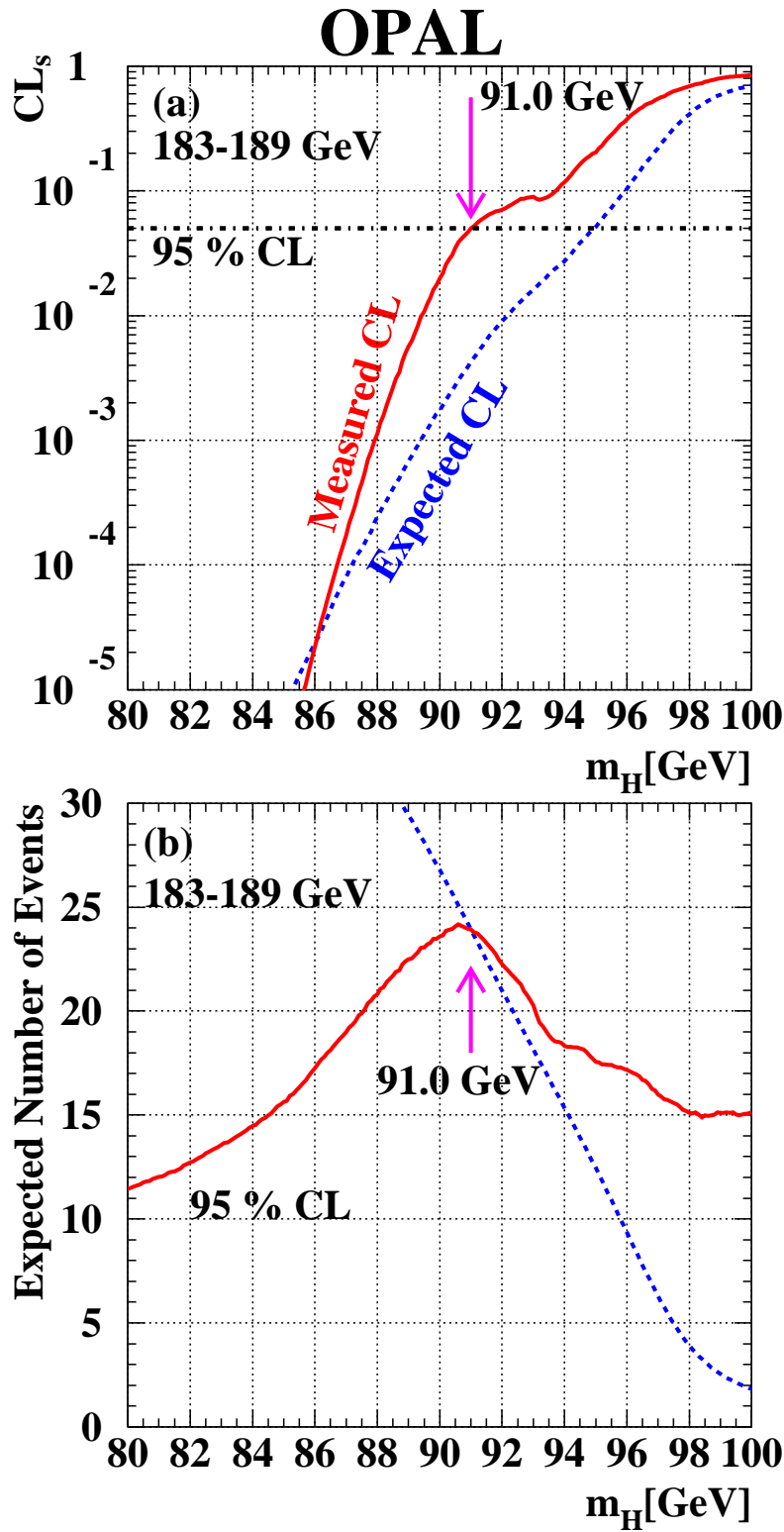,width=0.65\textwidth}}
\caption[]{\label{fig:smlimn95}\sl 
          Standard Model \Zo\Ho limits: (a) 
          The confidence level 
              for the signal hypothesis as observed
          (solid line) and
          expected on average for background-only experiments
          (dashed line), as functions of the SM Higgs boson mass. (b)
          The limit on the production rate for the SM Higgs boson at
          95\% CL (solid line) and the number of expected signal events
          (dashed line) as functions of the Higgs boson mass.
          The 95\% CL lower mass limit on the SM Higgs bosons is set at the
          point where the solid and dashed curve intersect: $m_{\mathrm H}>91.0$~GeV.
}
\end{figure}

To derive a lower bound on the mass of the SM Higgs boson, we
combine the results of the searches presented here with those of
our earlier searches at $\sqrts \approx 183$~GeV~\cite{pr183}.  OPAL's earlier
searches at energies of 172~GeV and lower have a negligible impact
on the expected limit because of the limited kinematic reach of the
lower-energy searches.
The systematic errors on the background expectations and on the signal selection
efficiencies have been
treated following the method given in Reference~\cite{ref:cousins}.
The confidence level for the signal hypothesis, $CL_s$, 
is shown in Figure~\ref{fig:smlimn95} (a).  It has been computed using
the weighted event-counting method
described in Section~5 of Reference~\cite{mssmpaper172}.

The number of expected signal events and the upper limit on the production
rate for signal events at the 95\% confidence level (CL) 
are given in Figure~\ref{fig:smlimn95} (b) as functions
of the Higgs boson mass hypothesis.
A lower mass bound of 91.0~GeV is obtained at the 95\%~CL,
while the average expected limit from a large number of fictitious
experiments assuming the background-only (zero signal) hypothesis is 94.9~GeV.
The probability to obtain a limit of 91.0~GeV or less in an ensemble of
background-only experiments is 4\%.

\begin{table}[htbp]
\begin{center}
\scalebox{0.94}{
\begin{tabular}{|c||c|c|c|c|c||c|c|} \hline
 $m_{\mathrm H}$ &\qq\Ho   &\nunu\Ho&\tautau\qq&\ee\Ho&\mm\Ho & 
\multicolumn{2}{|c|}{Expected signal} \\ \cline{7-8}
GeV&\Ho\ra\bb&      &         &      &      & 189~GeV & total \\
\hline\hline
 70&18.0 (14.96)&29.8 (8.75)&29.9 (3.56)&54.8 (2.65)&61.3
(2.81)&32.73&48.23  \\ \hline
 75&21.3 (15.38)&33.3 (8.52)&31.1 (3.24)&50.2 (2.11)&58.6
(2.35)&31.61&45.45  \\ \hline
 80&30.4 (18.28)&36.7 (7.87)&32.2 (2.83)&52.2 (1.84)&59.5
(1.99)&32.82&44.02  \\ \hline
 85&37.7 (17.91)&39.1 (6.67)&33.1 (2.31)&54.7 (1.53)&62.1
(1.65)&30.08&37.93  \\ \hline
 90&44.4 (14.86)&39.0 (4.78)&34.1 (1.69)&56.1 (1.12)&64.5
(1.22)&23.67&27.08  \\ \hline
 95&  47.0 ( 7.76)&35.4 (2.25)&34.3 (0.85)&55.0 (0.56)&64.9
(0.61)&12.03&12.56  \\ \hline
100&39.0 ( 0.99)&27.0 (0.38)&29.0 (0.11)&38.0 (0.08)&61.4 (0.09)& 1.65&
1.85  \\ \hline
%
%
\hline
Bkg.& 19.9 & 6.9 & 4.0 & 2.6 & 2.1 & 35.4 & 43.9 \\
$\sigma_{\mathrm{sys}}$& $\pm$3.0 & $\pm$0.8& $\pm$1.0 & 
$\pm$0.5 & $\pm$0.4 & $\pm$3.3 & $\pm$3.4 \\\hline\hline
Data   & 24 &  10  &  3 & 3 & 1 &  41 & 50 \\
\hline
\end{tabular}
}
\caption[]{\label{tab:smsummary}\sl
        Detection efficiencies in percent and numbers of expected SM Higgs
        boson events (in parentheses), at $\sqrts=189$~GeV,
        for each search channel separately,
        as a function of the Higgs boson mass.  
        The last two rows show the sum of expected SM backgrounds, the error
        on the background sum, and the numbers of candidates, for each
        channel separately.
        The last two columns show the total numbers of
        expected events in all channels for the present search at $\sqrts=189$~GeV,
        and also summed with the results obtained at 183~GeV~\cite{pr183}.
}
\end{center}
\end{table}

\subsection{Limits in the MSSM parameter space} 
\label{sect:mssmlimit}
The searches presented in this publication at $\sqrt{s}\approx 189$~GeV
are combined with previous OPAL Higgs 
searches~\cite{pr183,smpaper172,mssmpaper172,smhiggs91,mssmhiggs91,opalhiggs90} at 
$\sqrts$ between \mZ\ and 183~GeV.
We present 95\%~CL limits in the MSSM parameter space for the constrained MSSM with
six parameters, $m_0$, $M_2$, $A$, $\mu$, $\tanb=v_2/v_1$ and \mA\  in addition
to those of the SM. The definitions of these parameters can be found 
in~\cite{mssmpaper172}.
The consistency of the MSSM Higgs predictions
with our observed data has been calculated for
a large sample of models in the space spanned by these six parameters.  The scanning
strategy is described in detail in~\cite{mssmpaper172}.

\begin{figure}[p]
\centerline{\epsfig{file=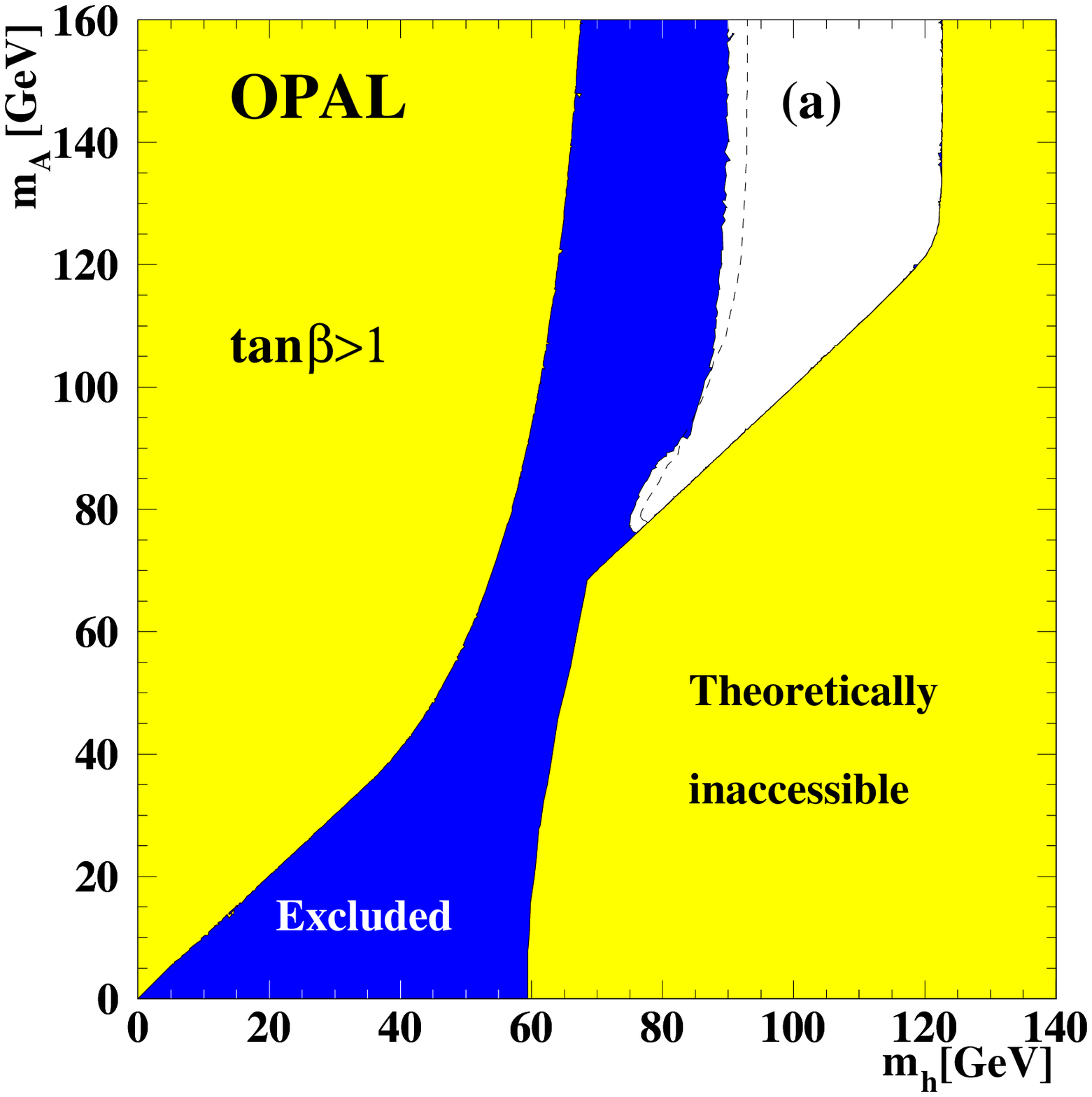,width=0.45\textwidth}\quad
            \epsfig{file=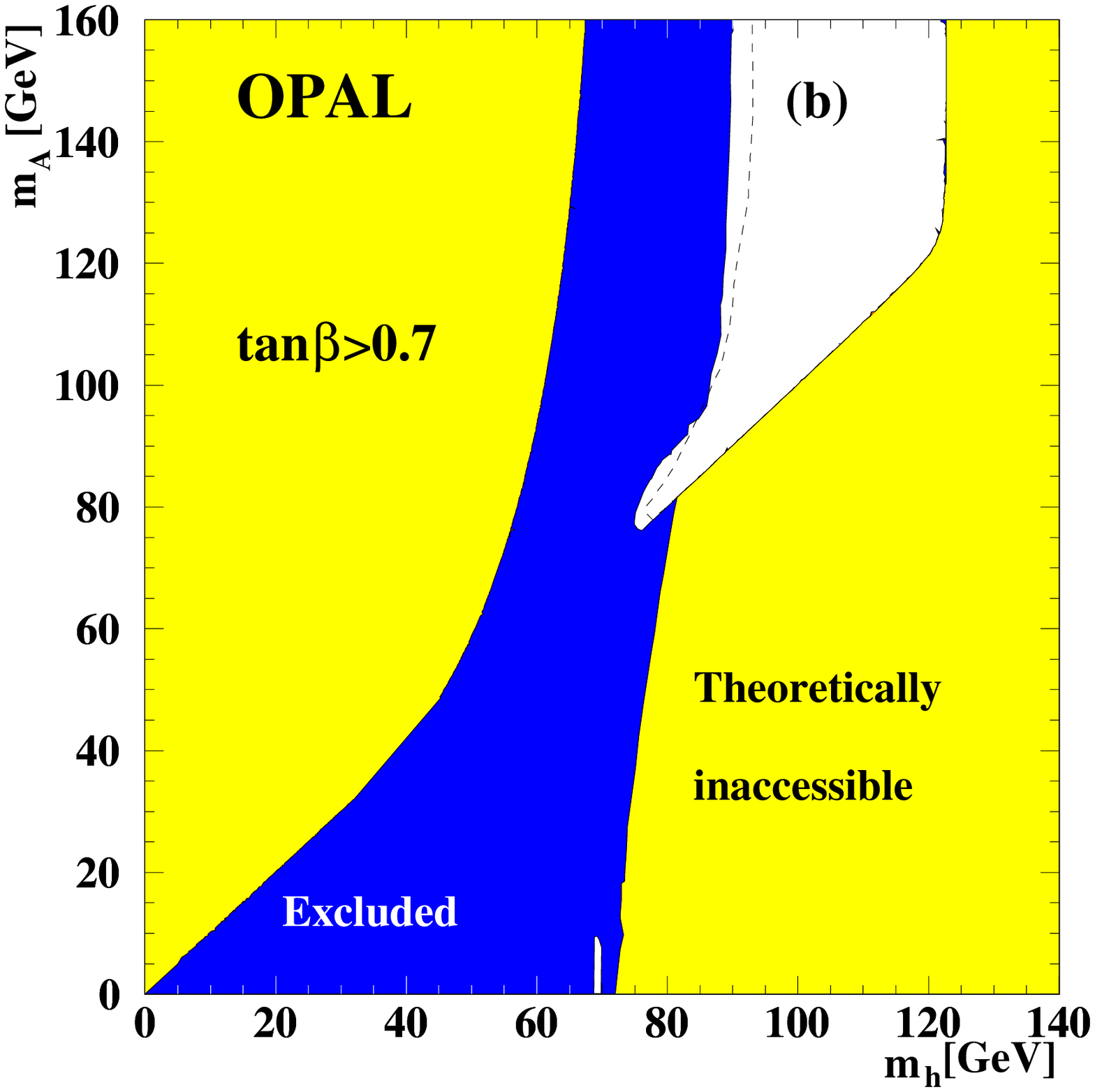,width=0.45\textwidth} }
\centerline{\epsfig{file=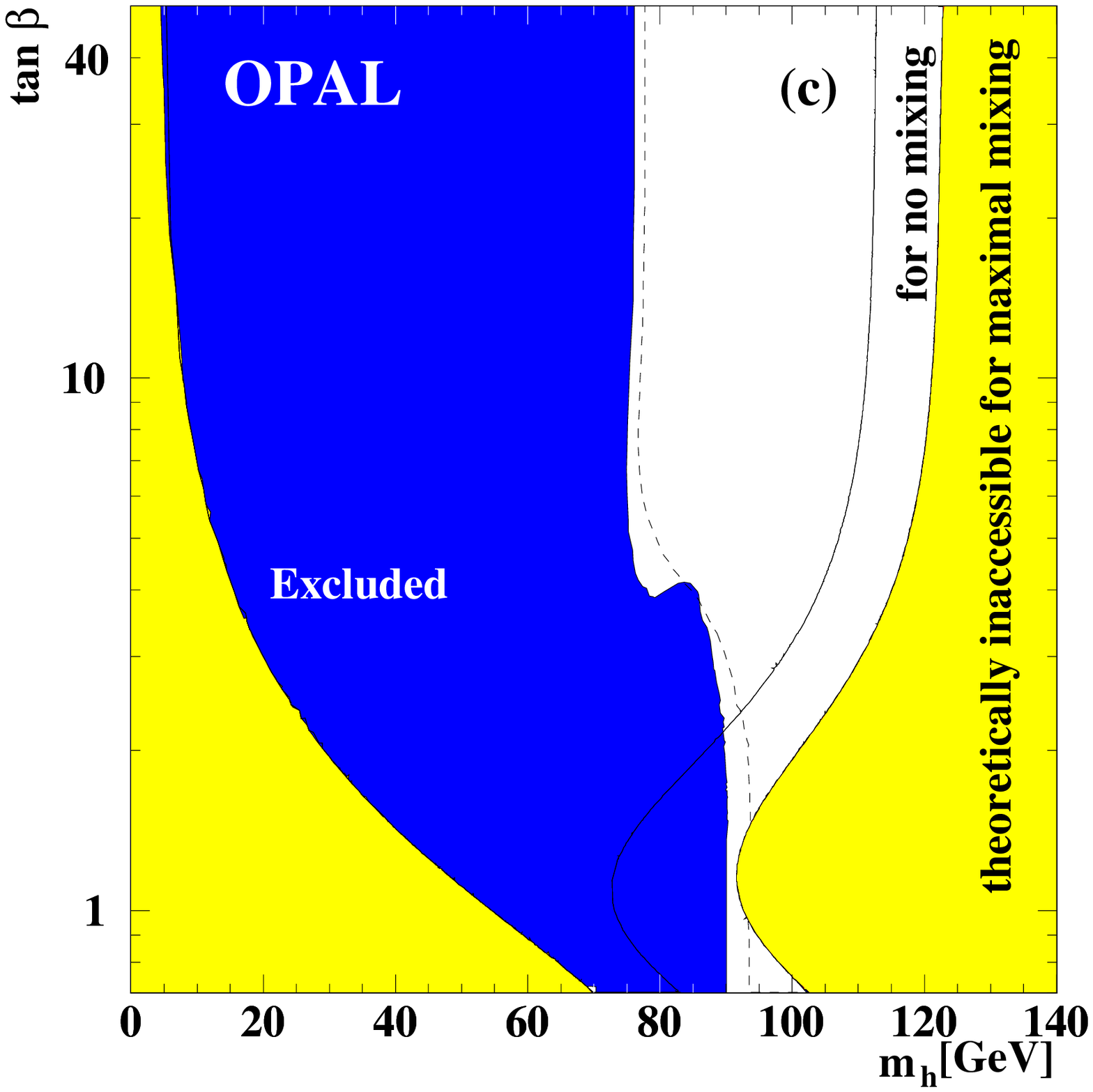,width=0.45\textwidth}\quad
            \epsfig{file=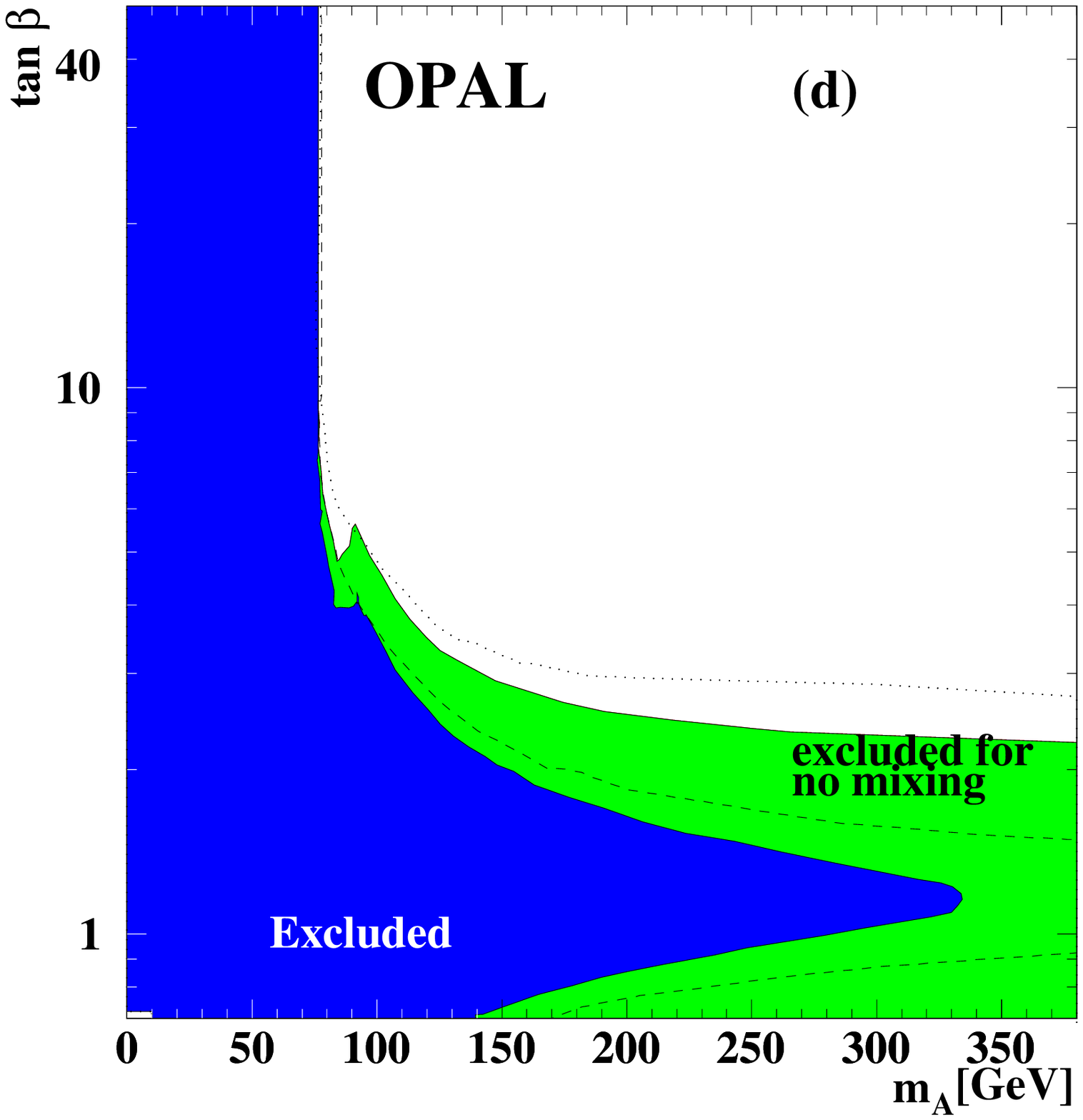,width=0.45\textwidth} }
\caption[]{\label{fig:scanA}\sl  
         The MSSM exclusion for the ``benchmark'' scan described in the text
         of Section~\ref{sect:mssmlimit}.
         Excluded regions are shown for
         (a) the $($\mh,~\mA$)$ plane for $\tanb>1$,
         (b) the $($\mh,~\mA$)$ plane for $\tanb>0.7$,
         (c) the $($\mh,~\tanb$)$ plane, and
         (d) the $($\mA,~\tanb$)$ plane.
         The black area is excluded at the 95\% CL.
         The grey areas in (a), (b) and (c) are theoretically inaccessible.
         In (b), because of the wider permitted range of $\tan\beta$, more
         $($\mh,~\mA$)$ points become available, and a small,
         unexcluded region appears at \mh$\approx$70~GeV, with
         \mA$<$10~GeV; it can also be seen in (c) and (d).
         Shown in (d) is the excluded region for no scalar top mixing.
         In all figures, the black region is excluded for all values of the
         scalar top mixing.  In (d), the grey area is
         excluded for the case of 
         no scalar top mixing.  The dashed lines indicate the boundary of the
         region expected to be excluded at the 95\% CL if only SM background
         processes are present.  The dotted line in (d) is the expected limit
         for the case with no scalar top mixing.
}
\end{figure}

Two ``benchmark'' situations (Scan~A of~\cite{mssmpaper172})
are first considered, corresponding to no
mixing and maximal mixing in the scalar-top sector
($A=0$ and $\sqrt{6}$~TeV, respectively), with $5<m_{\mathrm A}<2000$ GeV and
$0.7<\tan\beta<50$.  The remaining free parameters
are fixed to ``universal'' values:
the soft SUSY breaking mass parameters at the electroweak scale ({\it e.g.},
$m_{\mathrm Q}$, the ``left-up'' scalar quark mass)
are set to 1~TeV, $M_2$ is set\footnote{
In References~\cite{pr183} 
and~\cite{mssmpaper172} it was mistakenly stated
that $M_2$ was set to 1.0~TeV for Scan A.  The actual value of
$M_2$ used was 1.63~TeV in those papers.
}
to 1.63~TeV and $\mu$ is fixed to $-100$~GeV.
The top quark mass is fixed at 175 GeV.
The exclusions are
shown in Figure~\ref{fig:scanA} in the (\mh,\mA) plane,
the (\mh,\tanb)
 plane\footnote{The corresponding Figure 19(c) of 
Reference~\cite{pr183} mistakenly showed the 99\% CL exclusion contour.},
and the (\mA,\tanb)
plane.  Both cases of scalar top
mixing are included.  If for either choice of the mixing a parameter set is
not excluded, it is shown as unexcluded.

For $\tanb>1$, lower mass limits of $\mh>74.8$~GeV and $\mA>76.5$~GeV are
obtained, while 
the expected limits are $\mh>76.4$~GeV and $\mA>78.2$~GeV.
In the case of no scalar top mixing, we exclude the range $0.72<\tanb<2.19$,
while no \tanb\ is excluded in the case of maximal scalar top mixing.
For smaller $m_{\mathrm top}$ the excluded region of $\tan\beta$ becomes larger,
while for larger $m_{\mathrm top}$, it becomes smaller.
If $\tan\beta$ is allowed to vary between 0.7 and 1.0, the range of possiblilities
in the model is larger and a small, unexcluded region appears, visible in
Figures~\ref{fig:scanA} (b), (c) and (d), with \mh$\approx$70~GeV, and
\mA$<$10~GeV.  This region corresponds to models for which the dominant decay
of the \ho\ is into \Ao\Ao, while \Ao\ decays into \bb\ are kinematically
impossible, rendering the analyses of this paper inefficient.  Nearby regions
are excluded because the decay \ho\ra\bb\ proceeds, although with a reduced
branching fraction.

\begin{figure}[p]
\centerline{\epsfig{file=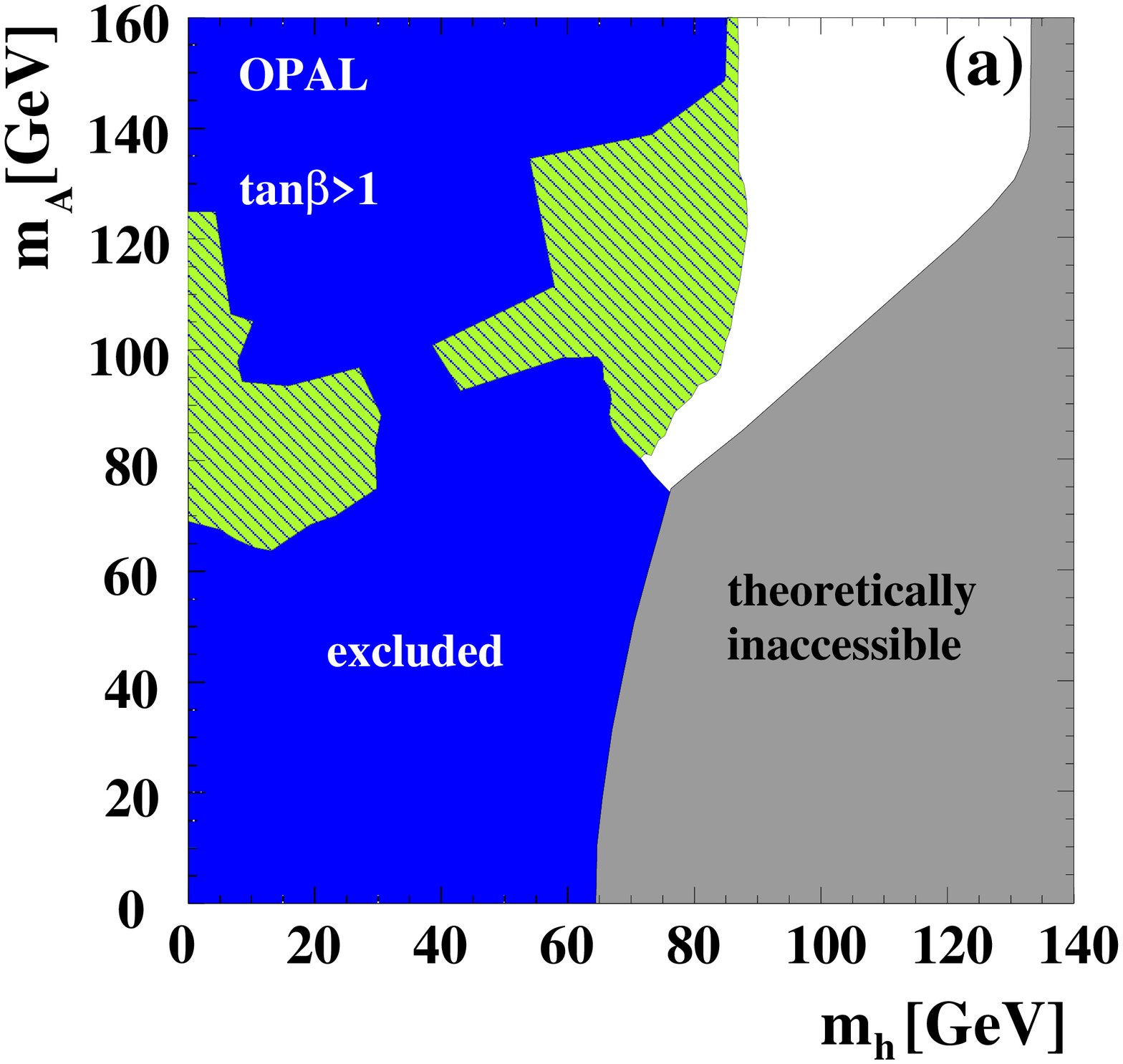,width=0.45\textwidth}\quad
            \epsfig{file=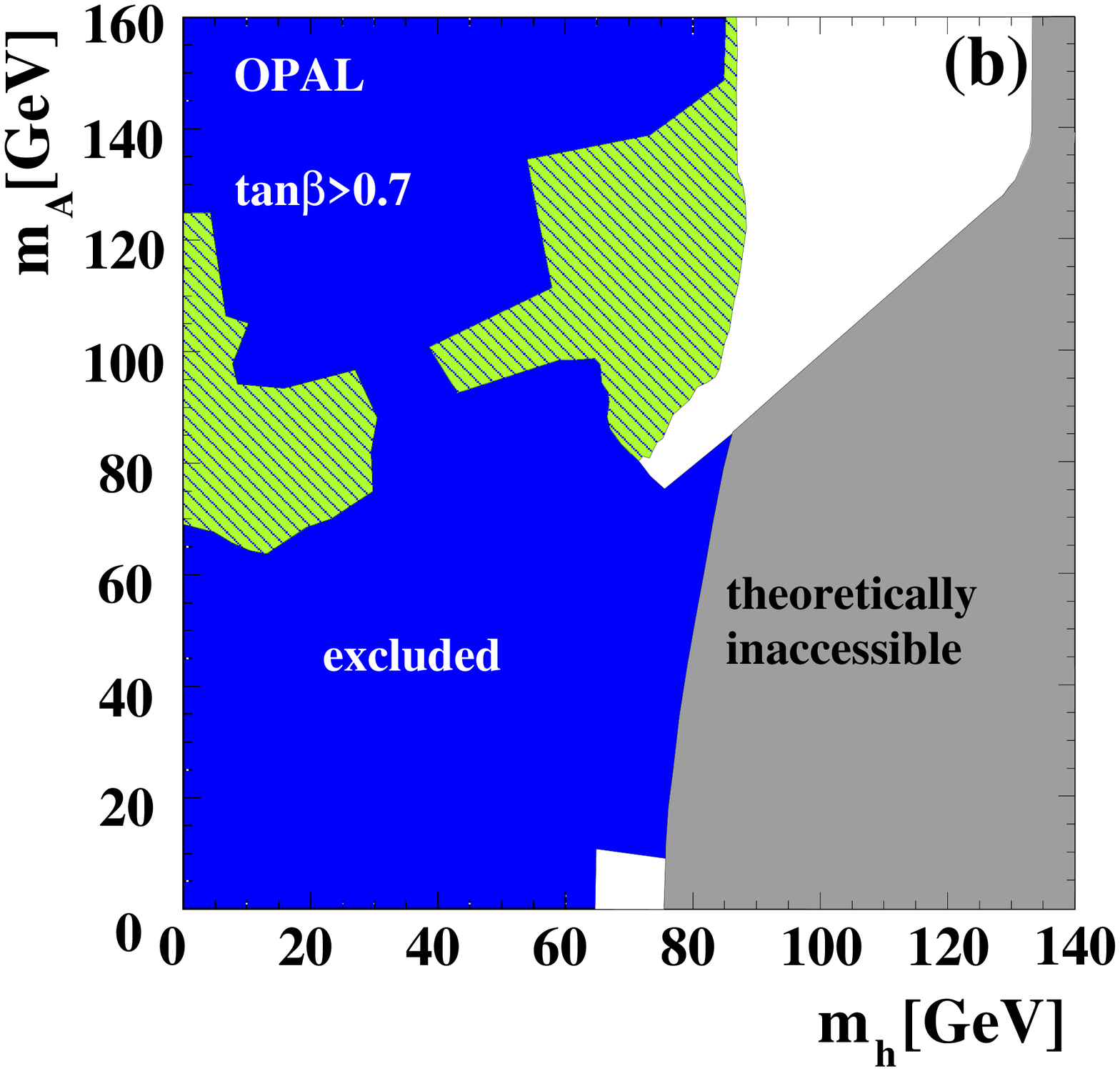,width=0.45\textwidth} }
\centerline{\epsfig{file=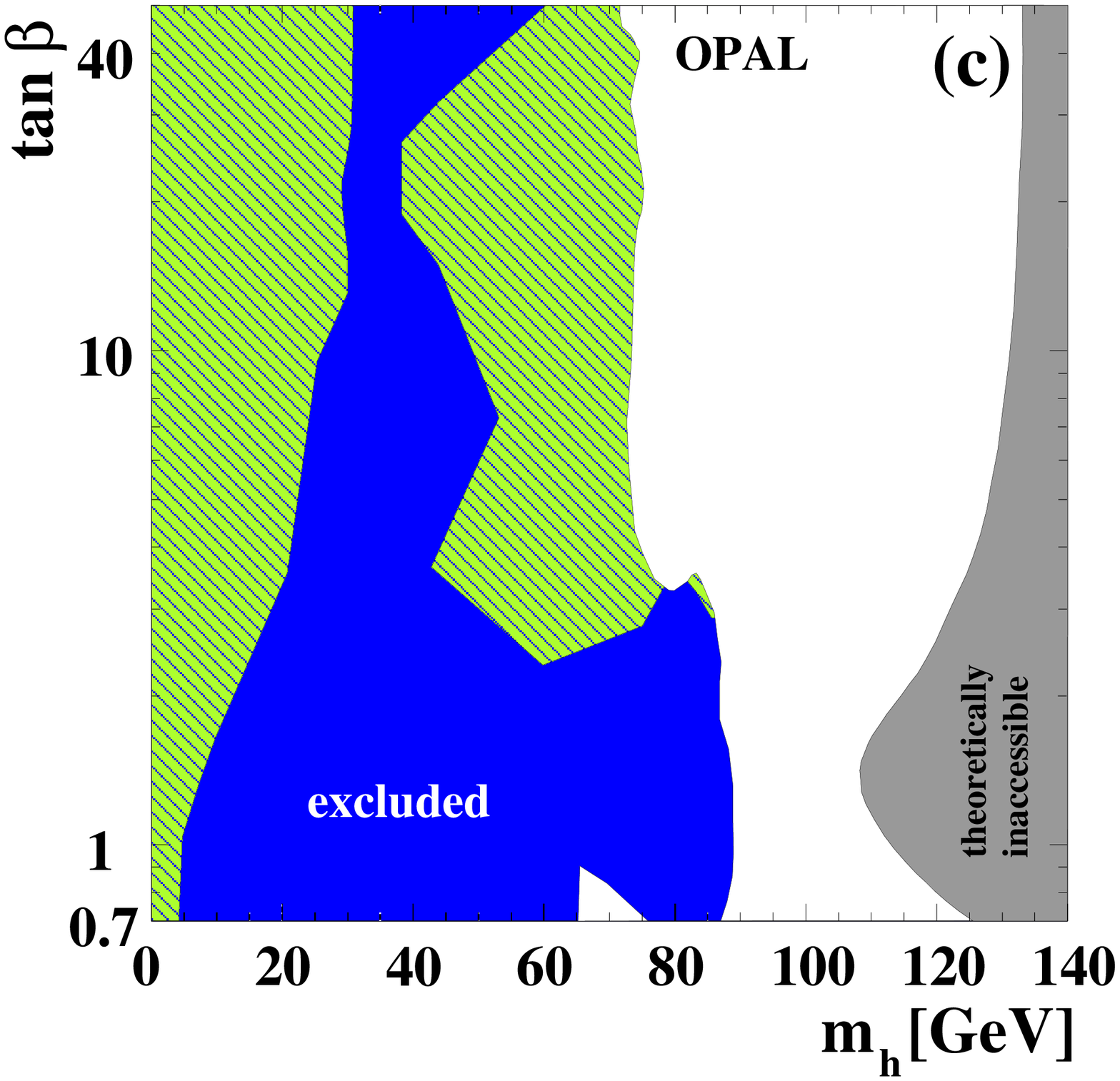,width=0.45\textwidth}\quad
            \epsfig{file=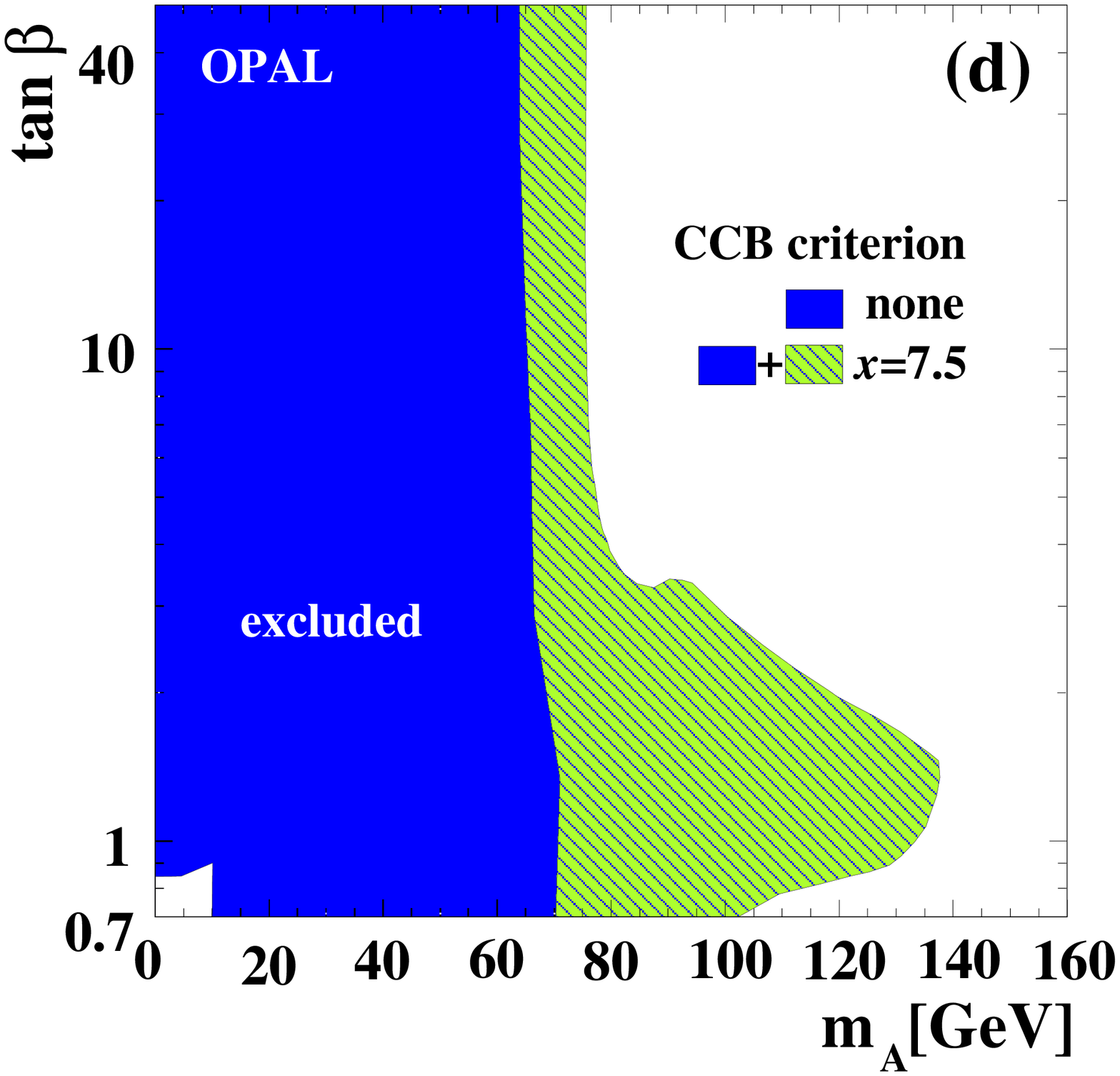,width=0.45\textwidth} }
\caption[]{\label{fig:scanC}\sl
         The MSSM exclusion for the general scan described in the text of
         Section~\ref{sect:mssmlimit}.
         Excluded regions are shown for
         (a) the $($\mh,~\mA$)$ plane for $\tanb>1$,
         (b) the $($\mh,~\mA$)$ plane for $\tanb>0.7$,
         (c) the $($\mh,~\tanb$)$ plane, and
         (d) the $($\mA,~\tanb$)$ plane.
         All exclusion limits are at the 95\% CL.
         The black areas are excluded without applying a CCB criterion (described
         in the text).
         The grey hatched areas are excluded when the CCB criterion is
         applied with $x=7.5$.
         The grey areas in (a), (b) and (c) are theoretically
         inaccessible.
}
\end{figure}

In a more general scan (Scan~C of~\cite{mssmpaper172}), all six parameters 
are varied independently within ranges motivated by theory, with the 
top quark mass
taken at 165, 175 and 185~GeV, which allows for about $\pm 2\sigma$ of the
current measurement error.  The ranges for the parameters of this scan
are $0<m_0<1$~TeV, $0<M_2<2$~TeV, $-2.5\cdot m_0<A<2.5\cdot m_0$, 
$-1<\mu<1$~TeV, $5<m_{\mathrm A}<2000$~GeV and $0.7<\tan\beta < 50$.
For some parameter sets, the heavier CP-even Higgs (\Ho ) is light
enough to be produced in the \Ho\Zo\ process, and in those cases 
the \ho\Zo\ searches are considered sensitive to it, thereby
extending the excluded area.
Parameter sets giving rise to chargino or neutralino
masses~\cite{neutralino183} or scalar top masses~\cite{lep2stop} already excluded by
OPAL searches are considered excluded here.  
Models which give rise to large \Zo\ra\ho\Zs\ or \Zo\ra\ho\Ao\ cross-sections
incompatible with the measured \Zo\ decay width (see~\cite{mssmpaper172})
are also considered excluded.
The results for this scan are shown in Figure~\ref{fig:scanC}.
The dark area is excluded at the 95\% CL.
The unexcluded region for \mh$\approx$70~GeV and \mA$<$10~GeV has become larger in this
more general scan, but it is still limited to the region of low $\tan\beta$.

For large values of $A$ and $\mu$, it is possible for the MSSM Lagrangian
to have minima which have non-zero vacuum expectation values for the $\tilde{\mathrm t}$
fields which break charge and colour symmetry~\cite{frere}.  One may
identify parameter sets for which the Lagrangian has such a minimum~\cite{casas}, but
the criteria for excluding such sets because they are inconsistent with
observation are substantially modified if it is possible
for the electroweak minimum to be only a local minimum and the tunneling rate to the
charge- and colour-breaking (CCB) minimum is longer than the age of the universe.
A specific model including
the effect of tunneling is available out of a number of different 
possibilities~\cite{kusenko}.  A simple, approximate criterion to avoid CCB minima
is~\cite{frere}
\begin{eqnarray*}
A^2 + 3\mu^2 < x(m_{\tilde{\mathrm{t}}_{\mathrm{L}}}^2 +
                 m_{\tilde{\mathrm{t}}_{\mathrm{R}}}^2 ),
\end{eqnarray*}
where \mstopL\ and \mstopR\ denote the left- and right-handed scalar top masses
and $x\approx$3.  For the specific calculation that includes the possibilities
of false vacua which tunnel only slowly to CCB vacua, this bound
is shown to be modified to $x\approx 7.5$.  We show in Figure~\ref{fig:scanC}
the regions in \mh, \mA and $\tan\beta$ in our general scan which
we do not exclude but which lead to Lagrangians with CCB minima according
to the looser criterion with $x=7.5$.
With this CCB criterion applied, absolute mass limits
\mA$>$76.0~GeV and \mh$>$72.2~GeV are derived for \tanb$>$1
at the 95\% CL.

\section{Conclusions} 

A search for neutral Higgs bosons has been performed based on the
data collected at $\sqrt{s} \approx 189$ GeV with an integrated
luminosity of approximately 170 pb$^{-1}$.
Searches have been performed for the Standard Model process
\ee\ra\Ho\Zo\  and the MSSM processes $\ee\ra\ho\Zo$, $\Ao\ho$.
The search channels are designed to detect 
\bb\ and \tautau\ decays of the Higgs bosons,
and the process \ho\ra\Ao\Ao\ 
is also considered.  No significant excess of candidates is observed
in the data beyond the expected Standard Model backgrounds, and we derive
the following limits at the 95\% confidence level.
For the SM Higgs boson, we obtain a lower mass bound of 91.0~GeV.
In the MSSM, we obtain the limits \mh$>74.8$~GeV and \mA$>76.5$~GeV
assuming \tanb$>$1, the mixing in the scalar top sector to be either 
zero or maximal, and the soft SUSY-breaking masses are 1~TeV.
In the general scan of MSSM parameters, excluding parameter sets which
result in Lagrangians with charge- and colour-breaking minima,
we derive absolute mass limits of
$\mh>72.2$~GeV and $\mA>76.0$~GeV for \tanb$>1$ at the 95\% confidence level.

\section*{Acknowledgements}
\par
We particularly wish to thank the SL Division for the efficient operation
of the LEP accelerator at all energies
 and for their continuing close cooperation with
our experimental group.  We thank our colleagues from CEA, DAPNIA/SPP,
CE-Saclay for their efforts over the years on the time-of-flight and trigger
systems which we continue to use.  In addition to the support staff at our own
institutions we are pleased to acknowledge the  \\
Department of Energy, USA, \\
National Science Foundation, USA, \\
Particle Physics and Astronomy Research Council, UK, \\
Natural Sciences and Engineering Research Council, Canada, \\
Israel Science Foundation, administered by the Israel
Academy of Science and Humanities, \\
Minerva Gesellschaft, \\
Benoziyo Center for High Energy Physics,\\
Japanese Ministry of Education, Science and Culture (the
Monbusho) and a grant under the Monbusho International
Science Research Program,\\
Japanese Society for the Promotion of Science (JSPS),\\
German Israeli Bi-national Science Foundation (GIF), \\
Bundesministerium f\"ur Bildung, Wissenschaft,
Forschung und Technologie, Germany, \\
National Research Council of Canada, \\
Research Corporation, USA,\\
Hungarian Foundation for Scientific Research, OTKA T-029328, 
T023793 and OTKA F-023259.\\


\clearpage\newpage

\clearpage\newpage

\clearpage\newpage

\clearpage\newpage

\clearpage\newpage

\clearpage\newpage

%

\end{document}